\documentclass[aps,prl,twocolumn,longbibliography,floatfix,10pt,superscriptaddress]{revtex4-2}

\usepackage{graphicx}% Include figure files
\usepackage{dcolumn}% Align table columns on decimal point
\usepackage{bm}% bold math
\usepackage{comment}
\usepackage{graphicx,epsf,bm,bbm,color,amsmath,amssymb,float, subfigure}

\usepackage{color}
\usepackage{textcomp}
\usepackage{dsfont}
\usepackage{tikz}
\usepackage{indentfirst}
\usepackage[colorlinks,citecolor=blue]{hyperref}
\usepackage{soul}
\usepackage{multirow}
\newcommand{\remove}[1]{}

\newcommand{\supplementarysection}{%
  \setcounter{figure}{0}% Reset figure counter
  \let\oldthefigure\thefigure% Capture figure numbering scheme
  \renewcommand{\thefigure}{S\oldthefigure}% Prefix figure number with S
  \setcounter{section}{0}% Reset figure counter
  \let\oldthesection\thesection% Capture section numbering scheme
  \renewcommand{\thesection}{S\oldthesection}% Prefix section number with S
  \setcounter{equation}{0}% Reset figure counter
  \let\oldtheequation\theequation% Capture section numbering scheme
  \renewcommand{\theequation}{S\oldtheequation}% Prefix section number with S
  \setcounter{table}{0}% Reset figure counter
  \let\oldthetable\thetable% Capture section numbering scheme
  \renewcommand{\thetable}{S\oldthetable}% Prefix section number with S
}

\newcommand{\bi}{\begin{itemize}}
\newcommand{\ei}{\end{itemize}}
\newcommand{\be}{\begin{enumerate}}
\newcommand{\ee}{\end{enumerate}}
\newenvironment{dfn}{{\vspace*{1ex} \noindent \bf Definition }}{\vspace*{1ex}}

\newcommand{\nn}{\nonumber}  %
\newcommand{\Fig}[1]{FIG.~\ref{#1}}
\newcommand{\Eq}[1]{Eq.~(\ref{#1})}

\newcommand{\ket}[1]{\left| #1 \right>} % for Dirac bras

	\newcommand{\beq}{\begin{eqnarray}}
	\newcommand{\eeq}{\end{eqnarray}}
	\newcommand{\bea}{\begin{eqnarray}\begin{aligned}}
	\newcommand{\eea}{\end{aligned}\end{eqnarray}}

\begin{document}

\title{Topological BF Theory construction of twisted dihedral quantum double phases from spontaneous symmetry breaking}

\author{Zhi-Qiang Gao}
\affiliation{Department of Physics, University of California, Berkeley, California 94720, USA}
\affiliation{Materials Sciences Division, Lawrence Berkeley National Laboratory, Berkeley, California 94720, USA}

\author{Chunxiao Liu}
\affiliation{Department of Physics, University of California, Berkeley, California 94720, USA}
\affiliation{Université Paris-Saclay, CNRS, Laboratoire de Physique des Solides, 91405 Orsay, France}

\author{Joel E. Moore}
\affiliation{Department of Physics, University of California,  Berkeley, California 94720, USA}
\affiliation{Materials Sciences Division, Lawrence Berkeley National Laboratory, Berkeley, California 94720, USA}


\begin{abstract}
Nonabelian topological orders host exotic anyons central to quantum computing, yet established realizations rely on case-by-case constructions that are often conceptually involved.
In this work, we present a systematic construction of nonabelian dihedral quantum double phases based on a continuous $O(2)$ gauge field. We first formulate a topological $S[O(2)\times O(2)]$ BF theory, and by identifying the Wilson loops and twist operators of this theory with anyons, we show that our topological BF theory reproduces the complete anyon data, and can incorporate all Dijkgraaf--Witten twists. Building on this correspondence, we present a microscopic model with $O(2)$ lattice gauge field coupled to Ising and rotor matter whose Higgsing yields the desired dihedral quantum double phase. A perturbative renormalization group analysis further suggests a direct transition from this phase to a $U(1)$ Coulomb or chiral topological phase at a stable multicritical point with emergent $O(3)$ symmetry. Our proposal offers an alternative route to nonabelian topological order with promising prospects in synthetic gauge field platforms.
\end{abstract}

\maketitle

\noindent\textit{Introduction.---} Topological orders are novel quantum phases of matter whose description lies beyond the traditional Landau paradigm of symmetry-breaking~\cite{Wen1989}. They are instead characterized by long-range entanglement~\cite{Wen2018} and emergent gauge fields~\cite{Wen1991,Read1991,Levin2005}, exhibiting unusual properties such as ground state degeneracy tied to the topology of space~\cite{Wen1989} and fractionalized anyonic quasiparticles with nontrivial statistics~\cite{TC}. These properties are universal and robust against arbitrary local perturbations~\cite{Niu1990}, making topological orders a natural platform for fault-tolerant quantum computation~\cite{QD}. Topological orders have been observed in fractional quantum Hall systems~\cite{Laughlin1983,Lopez1991,Zhang1992}; certain quantum spin and cold atom systems also offer promising routes to their realization~\cite{Anderson1973,Wen1991,Read1991}.

Among the topological orders, the \emph{nonabelian} ones play an essential role in realizing universal quantum computing~\cite{NayakRMP}, making the understanding and engineering of nonabelian topological orders a central task. Great efforts have been put into the design~\cite{Nat2023,Sala2025} of \emph{dihedral} quantum doubles (QDs), for which explicit lattice constructions exist, notably Kitaev’s QD Hamiltonians~\cite{QD} and Levin--Wen string-net models~\cite{Levin2005}. Yet they directly work with discrete degrees of freedom and the recipe to create a nonabelian gauge field is quite involved; they do not offer a systematic route to incorporate dihedral QDs into a broader hierarchy of gauge theories.

In this work we close that gap by showing that an $S[O(2)\times O(2)]$ gauge theory---a particular class of gauge theory endowed with a continuous $O(2)$ gauge field---reproduces the full anyon data of the dihedral QD with arbitrary \emph{Dijkgraaf--Witten twists} (DWTs)~\cite{DWT}. At the continuum level, we identify a topological BF-type action 
\beq
\mathcal{L}_\mathrm{BF} = \frac{in}{2\pi}a \mathrm{d}b,\quad (a,b)\sim (-a,-b),\label{eq:1}
\eeq
whose gauge-invariant observables map one-to-one onto the (worldlines of) anyons of the dihedral QD. We verify that the fractional statistics are
%(fusion rules, modular $S$ and $T$ matrices) 
identical to those of the dihedral QD without DWTs, and that all DWTs can be incorporated as attaching self Chern--Simons (CS) terms and introducing twisted boundary conditions to \Eq{eq:1}. This construction reveals how a continuous gauge theory belongs to the same universality class as the discrete dihedral QD. Such a correspondence inspires us to construct lattice models for dihedral QD phases whose local degrees of freedom are compact $O(2)$ rotors. We employ field theory analysis to map out the phase diagram, where the dihedral QD phases can naturally arise by condensing the $O(2)$ rotor. In addition, we find that this phase can undergo a direct transition to a $U(1)$ phase through a stable multicritical point with emergent $O(3)$ symmetry.

\noindent\textit{Quantum double and Dijkgraaf--Witten twist.---} The twisted quantum double $D^\omega(G)$ is a systematic construction of certain nonchiral topological orders \cite{Propitius1997,QD,Teo2023dihedraltwistliquid,supp}. The physical data of topological order is completely given by a finite group $G$ and a cohomology class $\omega \in H^3(G,U(1))$; the latter encodes the DWT. The physical data refers to the low energy gapped excitations (the anyons) and their statistics: an anyon $a \in \mathcal{A} = \{a,b,c,...\}$ is a shorthand for the corresponding excited state $|a\rangle$.  The statistics consists of the anyon quantum dimensions $d_a$, the fusion coefficients $N^c_{ab}$, the $F$-symbol $F^{abc}_d$ and the $R$-symbol $R^{ab}_c$: the quantum dimensions $\{d_a\}_{a \in \mathcal{A}}$ collectively determine the ground state degeneracy (GSD) of topological order on a genus $g$ surface: $\text{GSD} = \sum_{a \in \mathcal{A}} (d_a/|G|)^{2-2g}$, and any symbol $\mathcal{O}^{cd...}_{ab...}$ governs the transition amplitide for incoming anyons ($a,b,...$) and outcoming anyons ($c,d,...$).
For $D^\omega(G)$, the anyons are labeled by $a = \{(C\ell,\rho)\}$ where $C\ell$ runs over the conjugacy classes of $G$ and $\rho$ the irreducible representations (irreps) of the centralizer of any $g\in C\ell$, see Table \ref{table:odd} for odd $n$ (even $n$ case is left in \cite{supp}). The symbols $N^c_{ab}$, $F^{abc}_d$, and $R^{ab}_c$ can all be derived from the algebras of conjugacy classes and irreps of $G$ \footnote{The $F$ symbols for the non-twisted quantum double can be chosen to be trivial as this is the usual $D_n$ gauge theory in which all the pure gauge fluxes carry bosonic exchange statistics. For this reason, the $F$-symbols and $R$-symbols of $D^\omega(G)$ will only depend on the twist, i.e. the 3-cocycle $\omega\in H^3(G,U(1))$; and in particular they assume the form of $f^{g_1g_2g_3}_{g_4=g_1g_2g_3}$ and $R^{g_1g_2}_{g_3=g_1g_2}$, depending only on the group elements $g_i \in D_n$.}. The physical data for topological order can be equivalently determined from the modular $S$ and $T$ matrices \cite{Ng2023}.

\begin{table}[!htbp]
\begin{tabular}{cccccc}
\hline\hline
Anyon & Quantum & Number &Conjugacy & Centra- & \multirow{2}{*}{Irrep.} \\
Label&Dim.&of Anyons& Class & lizer &\\
\hline
$1$ & $1$ & $1$ & $\{1\}$ & $D_n$ & $\mathbf{1}_+$\\
$\eta$ & $1$ & $1$ & $\{1\}$ & $D_n$ & $\mathbf{1}_-$\\
$\psi_l$ & $2$ & $(n-1)/2$ & $\{1\}$ & $D_n$ & $\mathbf{2}_l$\\
$\psi_{l,r}$ & $2$ & $n(n-1)/2$ & $\{C_n^l,C_n^{n-l}\}$ & $\mathbb{Z}_n$ & $\mathbf{1}_r$\\
$\sigma_\pm$ & $n$ & $2$ & $\{M_1,...,M_n\}$ & $\mathbb{Z}_2$ & $\mathbf{1}_\pm$\\
\hline\hline
\end{tabular}
\caption{Anyon data of $D^\omega(D_n)$ for odd $n$. There are $(n^2+7)/2$ distinct anyons, and the total quantum dimension is $4n^2$. The anyons with quantum dimension equal to 2 are labeled by $\psi$'s, with $l=1,...,(n-1)/2$ and $r=-(n-1)/2,...,(n-1)/2$.}\label{table:odd}
\end{table}

\noindent\textit{$S[O(2)\times O(2)]$ BF theory for dihedral quantum double.---}
It is known that the $\mathbb{Z}_n$ QD can be described by a $U(1)\times U(1)$ BF theory \cite{Maldacena2001,Levin2005,Kou2008,Cho2011,Lu2016}. This construction is based on the fact that $\mathbb{Z}_n$ is a subgroup of $U(1)$. Moving to the non-abelian case of dihedral group, since $D_n=\mathbb{Z}_n\rtimes \mathbb{Z}_2$ and $O(2)=U(1)\rtimes \mathbb{Z}_2$, it is natural to also expect a $D_n$ QD phase to be described by $O(2)$ gauge fields. Concretely, we propose the following gauge theory:
\beq\label{DnL}
\mathcal{L}_{D_n}=\mathcal{L}_\mathrm{BF}+\frac{ik}{2\pi}a\mathrm{d}a,~~~ k=0,1,...,n-1.
% =\frac{ik}{2\pi}a\mathrm{d}a+\frac{in}{2\pi}a\mathrm{d}b
\eeq
This theory has the same appearance as the $U(1)$ BF gauge theory; however $a$ is an $O(2)$ gauge field arising from gauging the charge conjugation symmetry $\mathbb{Z}_2^C:~a\mapsto -a$ \cite{Moore1989,Amano1990}. The level of the self CS term, $2k$, guarantees that the theory is bosonic. 

The fate of field $b$ in $\mathcal{L}_\mathrm{BF}$ under $\mathbb{Z}_2^C$ needs special attention.
Crucially, when $a$ is promoted to $O(2)$, $b$ cannot remain invariant under $\mathbb{Z}_2^C$, since $\mathbb{Z}_2^a:~a\mapsto -a,b\mapsto b$ does not preserve \Eq{DnL}. Instead, the appropriate symmetry is
\beq\label{sym:Z2C}
\mathbb{Z}_2^C:~a\mapsto -a,~b\mapsto -b.
\eeq
This defines the theory \eqref{DnL} for the dihedral QD to be $S[O(2)\times O(2)]$ BF theory, with a gauge structure $S[O(2)\times O(2)]:=[O(2)\times O(2)]/\mathbb{Z}_2$ \footnote{There exists a similar but different non-Abelian generalization of the $U(1)\times U(1)$ BF theory. In the absence of DWT, note that the $U(1)\times U(1)$ BF theory has another $\mathbb{Z}_2$ symmetry which exchanges $a$ and $b$. In fact, it reveals the EM duality \cite{TC} in $D(\mathbb{Z}_n)$. Gauging this duality symmetry in the $U(1)\times U(1)$ BF theory, which has been extensively studied in the literature, can give rise to different twisted quantum liquid phases \cite{Barkeshli2012,Chen2017} and non-Abelian quantum Hall phases \cite{Barkeshli2010,Barkeshli2010L,Barkeshli2011}. However, this class of BF theory (which is called $U(1)\times U(1)\rtimes \mathbb{Z}_2$ theory) cannot capture the physics of the $D_n$ quantum doubles. In the absence of DWT, the $S[O(2)\times O(2)]$ BF theory still preserves this duality symmetry upon exchanging $a$ and $b$, which reflects the duality symmetry \cite{Ng2023} upon exchanging $\psi_{l,r}$ and $\psi_{l,-r}$ while preserving other anyons.}. For conciseness, in the rest of this letter we only present the case of odd $n$; the even $n$ case is treated in detail in \cite{supp}.

A subtle issue exists within this construction when considering DWTs: the self CS term, judging by its appearance in \Eq{DnL}, can only capture the \emph{even} half of the DWTs of $D^\omega(D_n)$, namely, those that have a trivial value in the $\mathbb{Z}_2$ factor of $H^3(D_n,U(1)) =\mathbb{Z}_{2n}=\mathbb{Z}_n\times \mathbb{Z}_2$. The rest with a nontrivial $\mathbb{Z}_2$ value are \emph{odd} DWTs originated from the nontrivial DWT in the charge conjugation $\mathbb{Z}_2^C$. As we will show below, these \emph{odd} DWTs are correctly reflected in the \emph{twisted} sectors of the Hilbert space of the theory \eqref{DnL}.

\noindent\textit{Correspondence between $D^\omega(D_n)$ and $S[O(2)\times O(2)]$ BF theory.}--- In order to show that $S[O(2)\times O(2)]$ BF theory is a valid description for the dihedral QD, we analyze the states and sectors of the $S[O(2)\times O(2)]$ BF theory on Riemann surfaces: the states are obtained from those of the $U(1)\times U(1)$ BF theory by gauging the symmetry $\mathbb{Z}_2^C$ defined in \eqref{sym:Z2C}. Consider a genus $g$ surface $\Sigma_g$ with noncontractible loops $\alpha_i$, $\beta_i$ with $i=1,2,...,g$, where $\alpha_i$ and $\beta_j$ intersect if and only if $i=j$. On $\Sigma_g$, the ground states of a $U(1)\times U(1)$ BF theory are  \cite{Barkeshli2010}
\beq \label{u1_pq_identification}
\bigotimes_{i=1}^g\ket{p_i, q_i},~~~ (p_i,q_i)\!\sim\! (p_i\!+\!n,q_i)\!\sim\! (p_i\!+\!2k,q_i\!+\!n),
\eeq
where $p_i,q_i \in \mathbb{Z}$ label zero modes (flat configurations) of $a$ and $b$, respectively.
To promote the $U(1)\times U(1)$ gauge structure to $S[O(2)\times O(2)]$, the $\mathbb{Z}_2^C$ symmetry needs to be further gauged, identifying 
\beq\label{o2_pq_identification}
(p_i, q_i) \sim (-p_i,- q_i).
\eeq
This gauging procedure introduces \emph{twisted} sectors in the Hilbert space: gauge fields $a$ and $b$ in a twisted sector change sign (i.e. antiperiodic) across at least one of the noncontractible loops. There are a total of $(2^{2g}-1)$ \emph{twisted} sectors \footnote{All the sectors (including the \emph{untwisted} one) are in one-to-one correspondence with the distinct flat connections on $\Sigma_g$ up to conjugation: $\mathrm{Hom}(\pi_1(\Sigma_g),G)/G$.}. A detailed counting (see \cite{supp}) shows that the number of flat configurations of the $S[O(2)\times O(2)]$ theory matches the GSD of $D^\omega(D_n)$.

A one-to-one correspondence between anyons of $D^\omega(D_n)$ and the operators of the $S[O(2)\times O(2)]$ BF theory can also be drawn. On a torus with $g=1$ \footnote{To show the correspondence it suffices to restrict the theory to a torus.}, depending on whether there are twists across of the gauge fields on the loops $\alpha$ and $\beta$, the total Hilbert space is resolved into four sectors $\mathcal{H}_{\bullet\bullet}$ with $\bullet\bullet=++$, $+-$, $-+$, and $--$. The untwisted sector $\mathcal{H}_{++}$ has its states labeled by $\ket{p,q}_{++}$, whereas each of the three twisted sectors contains only one state for odd $n$.

In the untwisted sector of $S[O(2)\times O(2)]$ BF theory, the operators can be straightforwardly identified with certain anyons of $D^\omega(D_n)$. Starting from the Wilson loop of the $U(1)\times U(1)$ BF theory 
\beq
w_{p,q}=\mathcal{P}\exp{\left(i\oint_\gamma pa+qb\right)},\label{eq:wzn}
\eeq
where $\mathcal{P}$ is path ordering and $\gamma$ is some closed loop on the torus. The ``electric charge'' $p$ and ``magnetic charge'' $q$ are subject to the identification \eqref{u1_pq_identification}. These Wilson loop operators satisfy Abelian fusion rules $w_{p,q}\times w_{p^\prime,q^\prime}=w_{p+p^\prime,q+q^\prime}$. The $\mathbb{Z}_2^C$ gauge invariant operators of the $S[O(2)\times O(2)]$ BF theory are 
\beq\label{eq:wdn}
\begin{aligned}
W_{p,q}&=w_{p,q}+w_{p,q}^\dagger, %=2\mathcal{P}\cos\left(\oint_\gamma pa+qb\right),
~(p,q)\neq (0,0),\\
W_{0,0}&= w_{0,0},
\end{aligned}
\eeq
subject to the additional equivalence relation \Eq{o2_pq_identification}, and we have $|p,q\rangle_{++} = W_{p,q}|0,0\rangle.$ 
The topological spin of $w_{p,q}$, $w^\dagger_{p,q}=w_{n+2k-p,n-q}$, and $W_{p,q}$ is identical. The number of inequivalent nontrivial operators in the BF theory, $(n^2-1)/2$, coincides with the number of anyons with quantum dimension 2 in $D^\omega(D_n)$, evidencing a one-to-one correspondence. The correspondence is further revealed by the fusion rule of $W_{p,q}$, given by
\beq\label{eq:wfusionodd}
W_{p,q}\times W_{p^\prime,q^\prime}=W_{p+p^\prime,q+q^\prime}+W_{p-p^\prime, q-q^\prime},
\eeq
for $(p,q)\neq (\pm p^\prime,\pm q^\prime)$. Therefore, these operators in $S[O(2)\times O(2)]$ BF theory all have a quantum dimension $d=2$, consistent with that of the $\psi$ anyons in $D^\omega(D_n)$. $W_{p,0}$ are identified with $\psi_{l=p}$
% ($\psi_{l=p}$ and $\bar{\psi}_{l=p}$) 
in Table \ref{table:odd}, while the operators $\{W_{p,q\neq 0}\}$ are identified with $\{\psi_{l,r}\}$. 
% The duality symmetry of exchanging $\psi_{l,r}$ and $\psi_{l,-r}$ corresponds to exchanging $p$ and $q$, or equivalently exchanging gauge fields $a$ and $b$.

The Wilson loop operators discussed above create states within the untwisted sectors. For the $S[O(2)\times O(2)]$ BF theory, there also exist twist operators creating states that live in the twisted sectors. Below we show that these twist operators can be identified with the remaining anyons of $D^\omega(D_n)$. This is achieved by matching the topological statistics of twist operators and anyons.

\begin{figure}[t]
  \centering
\includegraphics[width=0.95\linewidth]{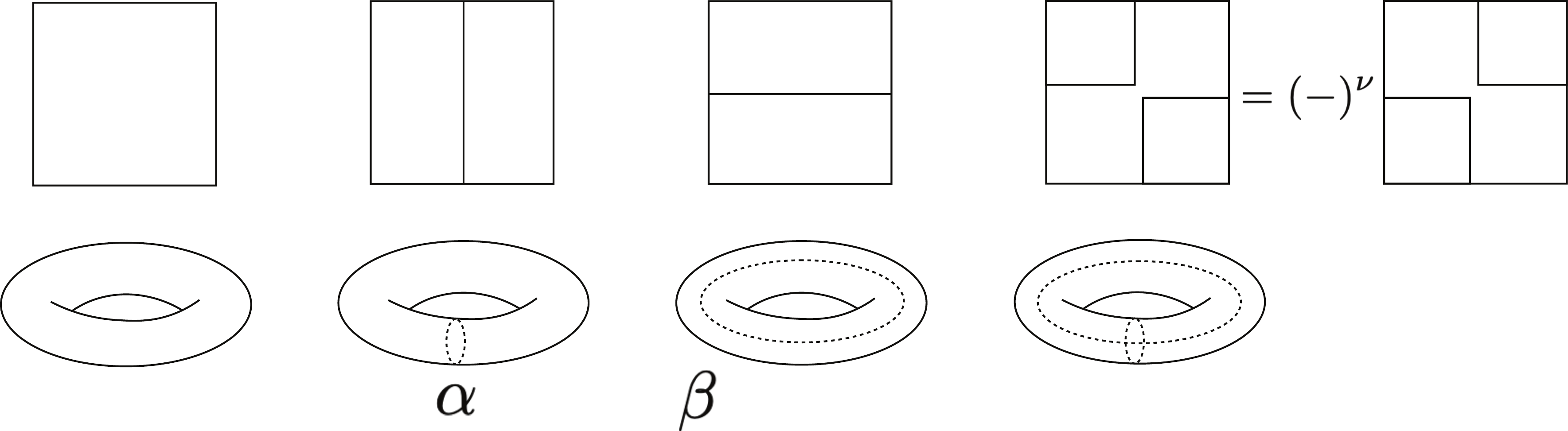}		\caption{$\mathbb{Z}_2$ sign of the $S$ and $T$ transformations.}\label{fig:ambiguity}
\end{figure}

To determine the topological spin and fusion statistics we calculate the $S$ and $T$ matrices for the twist operators, each of which creates the only state $\ket{0,0}_{+-/-+/--}$ in the corresponding twisted sector. Below we illustrate the procedure by calculating the topological spins, and the full analysis can be found in \cite{supp} (see also \cite{Verlinde1988,Dijkgraaf1989,Esko1993,Liu2013,Mong2022} for $S$ and $T$ transformations in $U(1)$ gauge theories). These transformations are illustrated in Fig.~\ref{fig:ambiguity} and read:
\bea
&\hat{S}\ket{0,0}_{\pm\mp}=\ket{0,0}_{\mp\pm},~\hat{S}\ket{0,0}_{--}=(-)^\nu\ket{0,0}_{--},~\\
&\hat{T}\ket{0,0}_{\pm\mp}=\ket{0,0}_{\pm -},~\hat{T}\ket{0,0}_{--}=(-)^\nu \ket{0,0}_{-+}.
\eea
Importantly, they contain a $\mathbb{Z}_2$ sign \cite{CFT} labeled by $\nu=0,1$, originating from whether the $\mathbb{Z}_2^C$ part of the gauge structure has a nontrivial DWT ($\nu=1$) or not ($\nu=0$) \cite{CFT,Barkeshli2013}. Here $\nu$ precisely captures the $\mathbb{Z}_2$ factor of $H^3(D_n,U(1)) = \mathbb{Z}_n\times \mathbb{Z}_2$. To see its effect, define
\bea
&\ket{1}=\frac{1}{\sqrt{2}}\left(\ket{0,0}_{++}+\ket{0,0}_{+-}\right),\\
&\ket{\eta}=\frac{1}{\sqrt{2}}\left(\ket{0,0}_{++}-\ket{0,0}_{+-}\right),\\
&\ket{\sigma_\pm}=\frac{1}{\sqrt{2}}\left(\ket{0,0}_{-+}\pm(-i)^\nu\ket{0,0}_{--}\right),
\eea
where a $T$ matrix calculation gives their topological spins $h_1=h_\eta=0$, $h_{\sigma_+}=\nu/4$, $h_{\sigma_-}=1/2+\nu/4$, identical to those of the $\eta$ and $\sigma_\pm$ anyons in Table.~\ref{table:odd}. We see that an effect of the $\mathbb{Z}_2$ sign $\nu$ is to change the topological spin of $\sigma_\pm$ by $1/4$ while preserving that of the other anyons. Physically, the nontrivial $\mathbb{Z}_2^C$ DWT corresponds to the gluing of a $\mathbb{Z}_2^C$ SPT state \cite{Barkeshli2013,Chen2017,Ryu2017,BBCW}, and is reflected in the additional signs of $F$-symbols \cite{Levin2005}.

\noindent\textit{Higgsing $O(2)$ gauge field on lattice.}--- The $S[O(2)\times O(2)]$ gauge theory described above suggests a natural route to realizing dihedral QDs from Higgsing an $O(2)$ gauge theory. Below, we introduce a lattice construction of the $D_n$ QD phase based on this Higgsing scenario. Unlike previous constructions based on Kitaev's proposal \cite{QD,Nat2023,Sala2025}, our construction is built in a physically transparent way, featuring a lattice $O(2)$ gauge field \cite{Jacobson2025} that can arise in nematic liquid crystals \cite{Liu2015,Liu2016}, cold atoms \cite{RMPsyn,Schweizer2019}, spin liquids \cite{Mao2020}, and certain quantum Hall systems \cite{Gomes2022}.

Consider the following classical model defined on 3D cubic lattice
\beq
H=H_{\mathrm{Maxwell}}+\sum_{\left<\mathbf{i}\mathbf{j}\right>}-J_s s_\mathbf{i}|\mathbf{u}_{\mathbf{i}\mathbf{j}}|s_\mathbf{j}-J_v\mathbf{v}_\mathbf{i} \cdot\mathbf{u}_{\mathbf{i}\mathbf{j}}\cdot\mathbf{v}_\mathbf{j}.\label{eq:LH}
\eeq
The matter fields defined on lattice site $\mathbf{i}$ are the two-component unit-norm real vector field $\mathbf{v}_\mathbf{i}$ and the Ising field $s_\mathbf{i}=\pm 1$. The $O(2,\mathbb{R})$ gauge field $\mathbf{u}_{\mathbf{i}\mathbf{j}}$ is a 2-by-2 real orthogonal matrix with determinant $|\mathbf{u}_{\mathbf{i}\mathbf{j}}|=\pm 1$ defined on the link $\left<\mathbf{i}\mathbf{j}\right>$. The Maxwell term of the $O(2)$ gauge field 
%\beq
%H_\mathrm{Maxwell}=-\frac{1}{2\kappa}\mathrm{tr}\sum_\square\prod_{\left<\mathbf{i}\mathbf{j}\right>\in\square}\mathbf{u}_{\mathbf{i}\mathbf{j}}\label{eq:Max}
%\eeq
in \Eq{eq:LH} is standard. The $O(2)$ gauge transformations are $s_\mathbf{i}\mapsto s_\mathbf{i}^\prime =|\mathbf{R}_\mathbf{i} |s_\mathbf{i}$, $  \mathbf{v}_\mathbf{i}\mapsto\mathbf{v}_\mathbf{i}^\prime =\mathbf{R}_\mathbf{i}\cdot\mathbf{v}_\mathbf{i}$, $\mathbf{u}_{\mathbf{i}\mathbf{j}}\mapsto\mathbf{u}^\prime_{\mathbf{i}\mathbf{j}}=\mathbf{R}_\mathbf{i}\cdot\mathbf{u}_{\mathbf{i}\mathbf{j}}\cdot\mathbf{R}^\mathbf{T}_\mathbf{j}$, where $\mathbf{R}_\mathbf{i}$ is the $O(2)$ gauge transformation matrix transforming under the two-dimensional irrep $\mathbf{2}_n$ of $O(2)$. The Ising field $s_\mathbf{i}$ and the vector field $\mathbf{v}_\mathbf{i}$ transform under the non-trivial one-dimensional irrep and the two-dimensional irrep $\mathbf{2}_n$ of $O(2)$, respectively. DWTs can also be realized in the model \Eq{eq:LH} by self CS terms. In the following we illustrate the phases using the model without DWT, and assume the continuous gauge fields are deconfined due to monopole suppression.

\begin{figure}[t]
  \centering
		\includegraphics[width=0.98\linewidth]{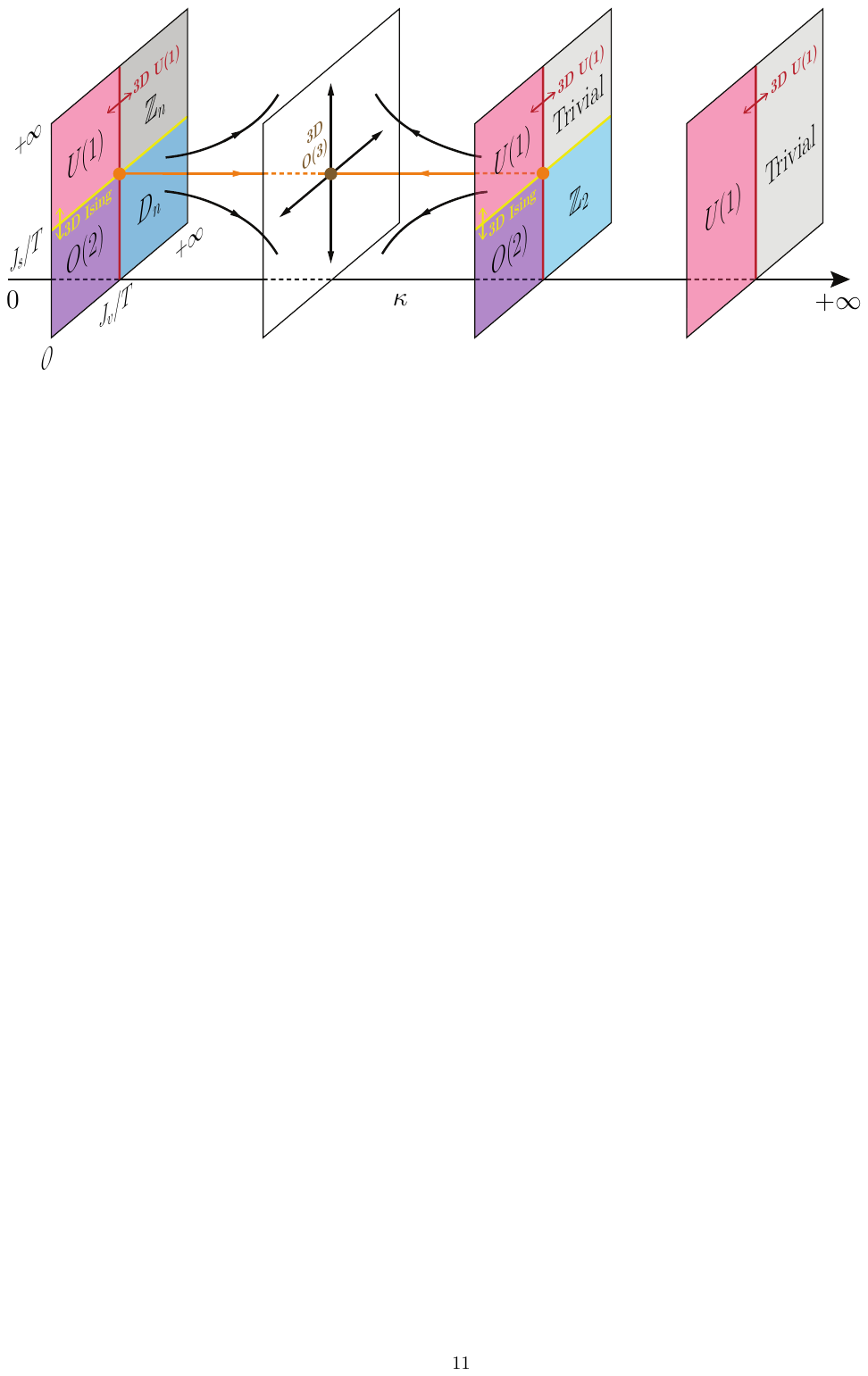}
		\caption{Schematic phase diagram of the lattice model \Eq{eq:LH} for $n>2$, with schematic RG flows derived from the effective field theory \Eq{eq:WF}. The relevant RG flow towards the $O(3)$ symmetric fixed point is highlighted in orange. The gauge structure of each phase is shown explicitly. The 3D $U(1)$ abelian Higgs and the 3D Ising transitions are marked in red and yellow, respectively. From left to right the three colored planes are phase diagrams for sufficiently small, mediate, and sufficiently large $\kappa$, respectively. The relative position of the $O(3)$ symmetric fixed point and the phase diagram for intermediate $\kappa$ is for illustrative purposes only and may differ from the actual situation.
        }\label{fig:schematic}
\end{figure}

The schematic phase diagram for theory \eqref{eq:LH} is shown in Fig.~\ref{fig:schematic}. For sufficiently small Maxwell coupling constant $\kappa$, all gauge fields are deconfined, and the condensation of $s_\mathbf{i}$ and $\mathbf{v}_\mathbf{i}$ can be separately tuned by $J_s/T$ and $J_v/T$ respectively, where $T$ is the temperature. This subdivides the phase region with deconfined gauge fields into four phases. Increasing $\kappa$ will confine the discrete gauge fields arising from properly Higgsing the $O(2)$ gauge field, as we analyze below.

When $J_v=0$, the $J_s$-term describes a $\mathbb{Z}_2$ gauge field interacting with an Ising matter \cite{Fradkin1979}. Increasing $J_s/T$ condenses the Ising field $s_\mathbf{i}$, thereby Higgsing the $\mathbb{Z}_2$ part of the $O(2)$ gauge field; the $U(1)$ part stays intact. This marks a 3D Ising transition \cite{Kogut1979,Fradkin1979} from an $O(2)$ Coulomb phase to a $U(1)$ Coulomb phase.

The same phase transition happens at $J_v=+\infty$ upon varying $J_s/T$. Sufficiently large $J_v$ condenses the vector field $\mathbf{v}_\mathbf{i}$, thereby Higgsing the $U(1)$ part of the $O(2)$ gauge field to $\mathbb{Z}_n$. Importantly, a residual $\mathbb{Z}_2\equiv O(2)/U(1)$ gauge redundancy still remains, which together with the unbroken $\mathbb{Z}_n$ gauge structure results in a $D_n$ QD phase. Here the Ising field $s_\mathbf{i}$ plays the role of $\eta$ anyon in $D(D_n)$ for odd $n$, and its condensation Higgses the $\mathbb{Z}_2$ part of the $D_n$ gauge field through a 3D Ising transition, leaving a $D(\mathbb{Z}_n)$ phase.

When $J_s=+\infty$, the $\mathbb{Z}_2$ part of the $O(2)$ gauge field is Higgsed through an Ising transition, and the theory \Eq{eq:LH} becomes a lattice Abelian Higgs model with charge-$n$ matter~\cite{Fisher1989,Dasgupta1981,Lee1985,Gao2022,Bonati2024}. Increasing $J_v/T$ condenses $\mathbf{v}_\mathbf{i}$, thereby Higgsing the $U(1)$ gauge structure to $\mathbb{Z}_n$; this marks a 3D $U(1)$ transition of an abelian Higgs model~\cite{Bonati2024} from the $U(1)$ Coulomb phase to the $D(\mathbb{Z}_n)$ phase~\cite{Dasgupta1981,Lee1985,Gao2022}. Similarly, when $s_\mathbf{i}$ is gapped at $J_s=0$, tuning $J_v/T$ also effects a 3D $U(1)$ transition between the $O(2)$ Coulomb phase and the $D(D_n)$ phase. A direct transition between $O(2)$ and $\mathbb{Z}_n$ phases can occur when the Ising and the $U(1)$ transitions meet with each other. The phase diagram is shown in the leftmost panel of Fig.~\ref{fig:schematic}.

\noindent\textit{Stable multicritical point.---} It is unnatural to expect a direct transition between $U(1)$ and $D_n$ phases, as neither one is the subgroup of the other. However, a Monte Carlo simulation (see End Matter for details) with monopole suppression suggests that for a wide range of $\kappa$ the multicritical point remains stable. Such a direct transition between the $D_n$ phase and the $U(1)$ Coulomb phase is similar to the deconfined quantum critical point and SPT transitions \cite{dqcp,Wang2017dqcp,Roberts2019,Senthil2024,Tsui2015,Tsui2017,Chatterjee2023,wu2024}. In what followings we propose a scenario based on a perturbative renormalization group (RG) analysis that a direct transition can happen at a stable multicritical point with an emergent $O(3)$ symmetry where the four stable phases meet.

In the model \eqref{eq:LH}, integrating out the $O(2)$ gauge field generates an effective coupling between the vector and the Ising fields. However, with monopole suppression and in absence of self CS terms, the $O(2)$ gauge field is gapless and the generated effective coupling is long-ranged~\footnote{We note that, with DWT, \textit{i.e.} self CS term in the $O(2)$ gauge field, the Coulomb phases become gapped chiral topological phases, in which integrating out the gauge field generates short-ranged effective coupling. Similarly, without monopole suppression, the continuous gauge fields are also gapped, resulting in short-ranged effective coupling as well}. In the RG analysis we adopt a simplified point effective coupling, and we conjecture that it extrapolates to the long-ranged case. The effective field theory reads
\beq
\mathcal{L}&=&|\partial_\mu\mathbf{\Phi}_v|^2+t_v|\mathbf{\Phi}_v|^2+r_v|\mathbf{\Phi}_v|^4+\tilde{\kappa}|\mathbf{\Phi}_v|^2\phi_s^2\nn\\
&\quad &+(\partial_\mu\phi_s)^2+t_s\phi_s^2+r_s\phi_s^4,\label{eq:WF}
\eeq
where $\mathbf{\Phi}_v$ and $\phi_s$ are the coarse-grained vector and Ising fields, respectively, $t_{v,s}\sim(J_{v,s}^c-J_{v,s})/T$ measures the deviation of the coupling constant $J_{v/s}$ from its critical value $J_{v,s}^c$, and $r_{v,s}>0$ denotes the self-interaction. The interaction between $\mathbf{\Phi}_v$ and $\phi_s$ generated by the $O(2)$ gauge field has coupling constant $\tilde{\kappa}\sim\kappa$. 

We first note that, when $\kappa$ is sufficiently small, the $O(2)$ gauge field is decoupled from the matter fields, and the two matter fields become decoupled in model \eqref{eq:LH}, resulting in a multicritical point where the $U(1)$ and the Ising transition lines, with $O(2)$ and $O(1)=\mathbb{Z}_2$ symmetry respectively, intersect with each other. The starting point of the perturbative RG analysis for \Eq{eq:WF} is this $O(2)\times O(1)$ symmetric point with $\tilde{\kappa}=0$ and $r_{v,s}>0$, at which parameters $t_v$, $t_s$, and $\tilde{\kappa}$ are relevant, while $r_v$ and $r_s$ are irrelevant. Among the relevant ones, $t_v$ and $t_s$ control the $U(1)$ and the Ising transitions, respectively, whereas $\tilde{\kappa}$ drives the RG trajectory towards a correlated fixed point conjectured to have an emergent $O(3)$ symmetry~\cite{Calabrese2003,PhysRevE.88.042141}, at which $\mathbf{\Phi}_v$ and $\phi_s$ combine into an $O(3)$ vector. Since $t_v$ and $t_s$ perturbations remain relevant in the parameter space of the lattice model, any RG trajectory---as long as it is not the direct trajectory from $O(2)\times O(1)$ to the $O(3)$ symmetric fixed point---will be driven away from the latter, as illustrated in the left two panels in Fig. \ref{fig:schematic}. This supports the stability of the multicritical point in the phase diagram. However, if the lattice model allows more relevant perturbations, the multicritical point will be no longer stable, as exemplified in another lattice construction using complex $O(2)$ gauge field given in \cite{supp}.

\noindent\textit{Confinement effects.---} Increasing of $\kappa$ also induces a two-step confinement of the gauge fields: the $\mathbb{Z}_n$ part of the gauge fields gets confined before the $\mathbb{Z}_2$ part for $n>2$ \cite{Kogut1980,Wu1982}. Thus, there exists an intermediate window of $\kappa$ where the phase diagram contains four phases, with $O(2)$, $U(1)$, $\mathbb{Z}_2$ and trivial gauge structures, respectively. The universality classes of the transitions among these phases remain unchanged as those in the small $\kappa$ case, and we conjecture that the multicritical point will still be stable. In addition, the transition between the $D(D_n)$ and the $D(\mathbb{Z}_2)$ phases and that between the $D(\mathbb{Z}_2)$ and the trivial phases are both described by the deconfinement to confinement transition of the $\mathbb{Z}_n$ gauge theory \cite{Kogut1979,Fradkin1979}. A direct transition between the $D(D_n)$ phase and the trivial phase is in principle allowed. When $\kappa$ is sufficiently large, the $\mathbb{Z}_2$ part of the gauge fields is further confined through a 3D Ising transition, leaving the $U(1)$ Coulomb and the trivial the only phases in the phase diagram. The multicritical point also disappears due to confinement of the Ising field $s_{\mathbf{i}}$. The phase diagram slices with intermediate and sufficiently large $\kappa$ are shown in the right two pannels in Fig. \ref{fig:schematic}.

\noindent\textit{Conclusion and outlook.---} In summary, we provide a new route to realizing nonabelian twisted dihedral quantum double phases $D^\omega(D_n)$ through continuous gauge fields: an $S[O(2)\times O(2)]$ BF theory that exactly reproduces the anyon content and modular data of $D^\omega(D_n)$, and a concrete Higgsing construction in an $O(2)$ lattice gauge model with matter fields that realizes these phases microscopically. By tracking Wilson loops and twist operators to anyons and matching their statistics, we establish a one-to-one correspondence that clarifies how discrete dihedral quantum double phases emerge from continuous gauge structure. In the lattice model, we find the dihedral quantum double phase can meet with a $U(1)$ Coulomb phase at a stable multicritical point with emergent $O(3)$ symmetry, enabling a direct continuous transition in analog to deconfined quantum criticality.

Our proposal for realizing nonabelian topological orders is based on continuous gauge theory which distinguishes ours from most existing proposals \cite{QD,Nat2023,Sala2025}. Our lattice construction allows us to access a large set of phases and several phase transitions (3D $U(1)$, 3D Ising, and $O(3)$ symmetric) by tuning a small set of parameters. Our construction can be systematized in the realization of more exotic nonabelian topological orders: for example, the point group quantum doubles via Higgsing $O(3)$ or $SU(2)$ gauge fields. Given the success in the experimental creation of synthetic nonabelian continuous gauge fields \cite{RMPsyn,Lin2011,Yang2019,Cheng2023,Liang2024,Madasu2025}, it is worthwhile to explore similar experiments realizing our proposed field theory \eqref{DnL}
and lattice model \eqref{eq:LH}.

\noindent\textit{Acknowledgments.---} ZQG acknowledges Yu-Ping Lin, Hui Yang, Ruihua Fan, Yan-Qi Wang, Salvatore Pace and Xiao-Gang Wen for helpful discussions. CL is grateful for valuable discussions with Nicolas Dupuis, Delamotte Bertrand, Jean-Noël Fuchs, Julien Vidal, Benoît Douçot, and Jean-Bernard Zuber on related topics. We thank Cenke Xu especially for helpful discussions on the lattice model and phase diagram. ZQG was funded through the U.S. Department of Energy, Office of Science, Office of Basic Energy Sciences, Materials Sciences and Engineering Division under Contract No. DE-AC02-05-CH11231 (Theory of Materials program KC2301). CL and JEM were supported by the U.S. Department of Energy, Office of Science, National Quantum Information Science Research Centers, Quantum Science Center. CL acknowledges support from the EPiQS program of the Gordon and Betty Moore Foundation.

\bibliographystyle{apsrev4-2}

\bibliography{o2.bib}

\appendix
\section{End Matter}

\begin{figure}[t]
  \centering
	\includegraphics[width=0.98\linewidth]{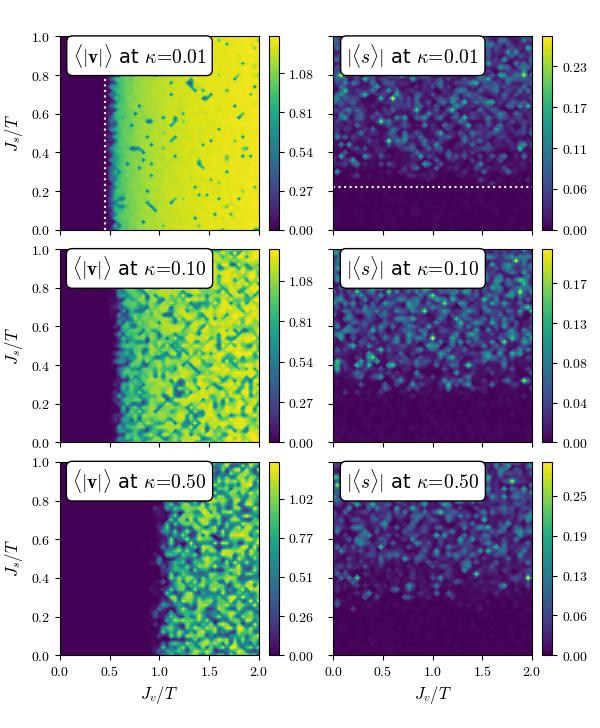}
	\caption{Numerical phase diagram in the $(J_v/T,J_s/T)$ parameter space for various $\kappa$ from Monte Carlo simulation.
        }\label{FIG:MC}
\end{figure}

Fig.~\ref{FIG:MC} shows the Monte Carlo simulation for the gauge theory \eqref{eq:LH} on $6\times 6\times 6$ cubic lattice with monopole suppression. Here $|\mathbf{v}|$ and $s$ are the order parameters for the $U(1)$ and $\mathbb{Z}_2$ symmetries, respectively. The vertical dashed line in the upper left panel denotes the critical value for the 3D XY transition, $J_v/T = 0.45422$, and the horizontal dashed line in the upper right panel denotes that for the 3D Ising transition, $J_s/T = 0.22165$. It can be seen that the multicritical point is always stable in the range of $\kappa$ we consider.

\clearpage

\onecolumngrid

\vspace{0.3cm}

\supplementarysection
 
\begin{center}
\Large{\bf Supplemental Material for ``Topological BF Theory construction of twisted dihedral quantum double phases from spontaneous symmetry breaking"}
\end{center}

\section{Data of the dihedral quantum double}

The fusion constants $N^c_{ab}$ for various anyon types can be deduced from Tables \ref{tab:n_odd_conj_cls} and \ref{tab:n_odd_irreps}.

The non-twisted quantum double has trivial fusion and braiding symbols due to the bosonic nature of the anyon statistics. The twisted quantum double models have $F$-symbols of $D^{[2k+\nu]}(D_n)$ that can be identified with the cocycle $\omega = [2k+\nu] \in H^3(D_n,U(1))$:
\begin{equation}
f^{g_1,g_2,g_3}_\omega = \omega(g_1,g_2,g_3)
=
e^{2\pi i (2k+\nu) \left(\frac{1}{n^2}(-1)^{s_2+s_3}r_1 \mathrm{Cell}_n\left[(-1)^{s_3}r_2+r_3\right] +\frac{1}{2}s_1s_2s_3\right)},
\end{equation}
where the function $\mathrm{Cell}_n[x]$ subtracts (or adds) integer multiples of $n$ to take $x$ back to the interval $(-(n-1)/2,(n-1)/2]$, and we denoted a generic element of $D_n$ by $g = C_n^rM^s$. The $R$-symbol is \cite{propitius1995topological,Teo2023dihedraltwistliquid}

\begin{equation}
r^{g_1g_2}_\omega = e^{\frac{2\pi i (2k+\nu)}{n^2} (a_2 \mathrm{Cell}_n[(-1)^{s_2}r_2+2s_1r_2] - s_1r_2^2) + \frac{i \pi (2k+\nu)}{2}s_1s_2},
\end{equation}  

For $n=2m$ even, the DWTs are captured by $H^3(D_n,U(1)) =\mathbb{Z}_n\times \mathbb{Z}_2^2$. Here we label the DWTs as $\omega=[k,\nu,\upsilon]$, with indices $\nu,\upsilon \in \{0,1\}$. We term the ones with $\nu=\upsilon=0$ as \emph{regular}, and the rest \emph{anomalous}. $F$-symbols of $D^{[k,\nu,\upsilon]}(D_{2m})$ can be identified with the cocycle $\omega = [k,\nu,\upsilon] \in H^3(D_{2m},U(1))$:
% \cite{propitius1995topological,Teo2023dihedraltwistliquid}:

\begin{equation}
f^{g_1,g_2,g_3} = \omega(g_1,g_2,g_3)
=
e^{2\pi i \left(\frac{k}{n^2}(-1)^{s_2+s_3}r_1 \mathrm{Cell}\left[(-1)^{s_3}r_2+r_3\right] +\frac{\nu}{2}s_1s_2s_3+\frac{\nu+\upsilon}{2}r_1s_2s_3\right)},
\end{equation}

And the $R$-symbol is \cite{Teo2023dihedraltwistliquid}
\begin{equation}
r^{g_1g_2}_\omega = \left\{\begin{array}{ll}
e^{-\frac{i \pi k}{n}r_2 + \frac{i \pi}{2} (\nu+ \upsilon)\mathrm{Cell}_2[m] s_2},& \text{when }(r_1,s_1) = (m,0),\\
e^{\frac{2\pi ik}{n^2} (a_2 \mathrm{Cell}_n[(-1)^{s_2}r_2+2s_1r_2] - s_1r_2^2) + \frac{i \pi }{2}(\nu s_1s_2 + (\nu+\upsilon)\mathrm{Cell}_2[r_1]s_2)}, &\text{otherwise.}\end{array}\right.
\end{equation}

\section{Equivalence between $S[O(2)\times O(2)]$ BF theory and $D_n$ quantum double}

\subsection{Ground State Degeneracy and Quantum Dimension}

This section provides a general counting of the ground state degeneracy on genus-$g$ Riemann surfaces, which dictates the anyon quantum dimensions of the theory. However, to extract fusion and braiding, one needs to explicitly calculate the modular $S$ and $T$ matrices, which will be left for the next section.

\begin{figure}[t]
  \centering
		\includegraphics[width=0.5\linewidth]{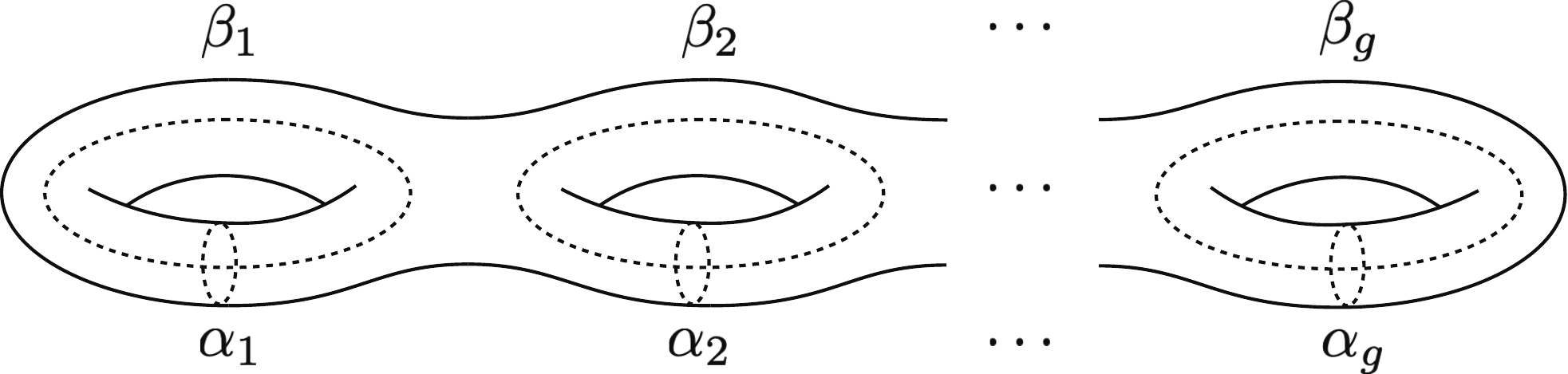}
		\caption{Genus-$g$ Riemann surface. Homology cycles $\alpha_i,~\beta_i$ with $i=1,2,...,g$ are marked by dashed loops, where $\alpha_i$ and $\beta_j$ intersect if and only if $i=j$.}\label{fig:genus}
\end{figure}

Gauging $\mathbb{Z}_2^C$ symmetry to promote the $U(1)\times U(1)$ BF theory to $S[O(2)\times O(2)]$ introduces \emph{twisted} sectors in the Hilbert space. Consider a genus-$g$ Riemann surface $\Sigma_g$ with inequivalent homology cycles (non-contractible loops) $\alpha_i,~\beta_i$ with $i=1,2,...,g$, where $\alpha_i$ and $\beta_j$ intersect if and only if $i=j$ (see \Fig{fig:genus} for illustration). A \emph{twisted} sector of the Hilbert space is defined as the states on which both the gauge fields $a$ and $b$ change sign (i.e. antiperiodic) on at least one of the cycles. All the sectors (including the \emph{untwisted} one) are in one-to-one correspondence with the distinct flat connections on $\Sigma_g$ up to conjugation: $\mathrm{Hom}(\pi_1(\Sigma_g),G)/G$. Therefore, there are a total of $(2^{2g}-1)$ \emph{twisted} sectors.

\begin{table}
\centering
\begin{tabular}{c|ccc}
\hline
\hline
Conj. Cls. of&       \multirow{2}{*}{$1$} &  \multirow{2}{*}{$C^i$} & \multirow{2}{*}{$M$} \\
$D_n$ ($n$ odd) &&&\\
\hline
$1$   & $1$ &  $C^i$ & $M$  \\
$C^j$ & $C^j$& $C^{i+j}\oplus C^{i-j}$ & $2M$\\
$M$ &  $M$  & $2M$ & $1\oplus 2\left(\bigoplus\limits_{k=1}^{(n-1)/2} C^k\right)$\\
\hline
\hline
\end{tabular}
\caption{Fusion rule for the conjugacy classes of $D_n$ for odd $n$. Here $i-j$ and $i+j$ are understood in the mod-$n$ sense. $i,j=1,2,...,(n-1)/2$.}\label{tab:n_odd_conj_cls}
\end{table}

\begin{table}
\centering
\begin{tabular}{c|ccc}
\hline
\hline
Irreps of &       \multirow{2}{*}{$\mathbf{1}_+$} &  \multirow{2}{*}{$\mathbf{1}_-$} & \multirow{2}{*}{$\mathbf{2}_i$} \\
$D_n$ ($n$ odd) &&&\\
\hline
$\mathbf{1}_+$   & $\mathbf{1}_+$ &  $\mathbf{1}_-$ & $\mathbf{2}_i$   \\
$\mathbf{1}_-$ & $\mathbf{1}_-$& $\mathbf{1}_+$ & $\mathbf{2}_i$\\
$\mathbf{2}_j$  & $\mathbf{2}_j$  & $\mathbf{2}_j$ & $\left\{\begin{array}{ll} \mathbf{2}_{i-j}\oplus \mathbf{2}_{i+j}, & i\neq j \\ \mathbf{1}_+ \oplus \mathbf{1}_- \oplus \mathbf{2}_{2j}, & i=j\end{array}\right.$\\
\hline
\hline
\end{tabular}
\caption{Fusion rule for the irreps of $D_n$ for odd $n$. Here $i-j$ and $i+j$ are understood in the mod-$n$ sense. $i,j=1,2,...,(n-1)/2$.}\label{tab:n_odd_irreps}
\end{table}

\begin{table}
\centering
\begin{tabular}{c|ccccc}
\hline
\hline
{Conj. Cls.}&        \multirow{2}{*}{$1$} &   \multirow{2}{*}{$C^i$} &  \multirow{2}{*}{$C^m$} &  \multirow{2}{*}{$M$} &  \multirow{2}{*}{$M'$}\\
of $D_{2m}$&&&&&\\
\hline
$1$  &  $1$ &  $C^i$ & $C^m$ & $M$ & $M'$\\
$C^j$  & $C^j$ &  $\left\{\begin{array}{ll}C^{i-j}\oplus C^{i+j},&i+j \neq m\\ C^{i-j}\oplus 2C^m, &i+j=m\end{array}\right.$ & $C^{m-i}\oplus C^{m+i}$ & $\left\{\begin{array}{ll}2 M, & j\text{ odd}\\
2M', & j \text{ even}\end{array}\right.$ & $\left\{\begin{array}{ll}2 M', & j\text{ odd}\\
2M, & j \text{ even}\end{array}\right.$\\
$C^m$ & $C^m$ & $C^{m-j}\oplus C^{m+j}$ & $1\oplus 1$ & $\left\{\begin{array}{ll}2 M, & m\text{ odd}\\
2M', & m \text{ even}\end{array}\right.$ & $\left\{\begin{array}{ll}2 M', & m\text{ odd}\\
2M, & m \text{ even}\end{array}\right.$\\
$M$ & $M$ & $\left\{\begin{array}{ll}2 M, & i\text{ odd}\\
2M', & i \text{ even}\end{array}\right.$ & $\left\{\begin{array}{ll}2 M, & m\text{ odd}\\
2M', & m \text{ even}\end{array}\right.$ & $1\oplus C^m \oplus 2 \left(\bigoplus\limits_{k=1,k\text{ even}}^{m-1} C^k\right)$ & $2 \left(\bigoplus\limits_{k=1,k\text{ odd}}^{m-1} C^k\right) $ \\
$M'$ & $M'$ & $\left\{\begin{array}{ll}2 M', & i\text{ odd}\\
2M, & i \text{ even}\end{array}\right.$ & $\left\{\begin{array}{ll}2 M, & m\text{ odd}\\
2M', & m \text{ even}\end{array}\right.$ & $2\left( \bigoplus\limits_{k=1,k\text{ odd}}^{m-1} C^k\right) $ &  $1\oplus C^m \oplus 2 \left(\bigoplus\limits_{k=1,k\text{ even}}^{m-1} C^k\right)$  \\
\hline
\hline
\end{tabular}
\caption{Fusion rule for the conjugacy classes of $D_n$ for even $n=2m$. Here $i-j$ and $i+j$ are understood in the mod-$n$ sense. $i,j=1,2,...,m-1$.}\label{tab:n_even_conj_cls}
\end{table}

\begin{table}
\centering
\begin{tabular}{c|ccccc}
\hline
\hline
{Irreps of}&        \multirow{2}{*}{$\mathbf{1}_1$} &   \multirow{2}{*}{$\mathbf{1}_e$} &  \multirow{2}{*}{$\mathbf{1}_m$} &  \multirow{2}{*}{$\mathbf{1}_f$} &  \multirow{2}{*}{$\mathbf{2}_i$}\\
$D_{2m}$ &&&&&\\
\hline
$\mathbf{1}_1$ & $\mathbf{1}_1$ & $\mathbf{1}_e$ & $\mathbf{1}_m$ & $\mathbf{1}_f$ & $\mathbf{2}_i$ \\
$\mathbf{1}_e$ & $\mathbf{1}_e$ & $\mathbf{1}_1$ & $\mathbf{1}_f$ & $\mathbf{1}_m$ & $\mathbf{2}_{m-i}$\\ 
$\mathbf{1}_m$ & $\mathbf{1}_m$ & $\mathbf{1}_f$ & $\mathbf{1}_1$ & $\mathbf{1}_e$ & $\mathbf{2}_i$ \\
$\mathbf{1}_f$ & $\mathbf{1}_f$ & $\mathbf{1}_m$ & $\mathbf{1}_e$ & $\mathbf{1}_1$ & $\mathbf{2}_{m-i}$\\
$\mathbf{2}_j$ & $\mathbf{2}_j$ & $\mathbf{2}_{m-j}$ & $\mathbf{2}_j$ & $\mathbf{2}_{m-j}$ & $\left\{\begin{array}{ll} \mathbf{2}_{i-j}\oplus \mathbf{2}_{i+j}, & i\neq j\text{ and }i+j \neq m,\\
\mathbf{1}_e \oplus \mathbf{1}_f, & i+j = m,\\
\mathbf{1}_1\oplus \mathbf{1}_m, & i=j\text{ and }2j\neq m,\\
\mathbf{1}_1\oplus \mathbf{1}_e\oplus \mathbf{1}_m \oplus \mathbf{1}_f, & i=j\text{ and }2j=m.\end{array}\right.$\\
\hline
\hline
\end{tabular}
\caption{Fusion rule for the irreps of $D_n$ for even $n=2m$. Here $i-j$ and $i+j$ are understood in the mod-$n$ sense. $i,j=1,2,...,m-1$.}\label{tab:n_even_irreps}
\end{table}

On $\Sigma_g$, the ground states of a $U(1)\times U(1)$ BF theory are
\beq
\bigotimes_{i=1}^g\ket{p_i, q_i},
\eeq
where $p_i,q_i \in \mathbb{Z}$ label zero modes of $a$ and $b$, respectively, subject to the identification
\beq \label{idodd}
(p,q)\sim (p+n,q)\sim (p+2k,q+n).
\eeq
The nontrivial element of the $\mathbb{Z}_2^C$ symmetry acts on these states as
\beq
\bigotimes_{i=1}^g\ket{p_i, q_i}\mapsto\bigotimes_{i=1}^g\ket{-p_i,- q_i}.
\eeq

Below we discuss the cases of odd $n$ and even $n$ separately.

\subsubsection{Odd $n$}

For odd $n$, the state $\otimes_i\ket{p_i=0,q_i=0}$ is automatically invariant under the $\mathbb{Z}_2^C$ action. To find the other invariant states, consider the states symmetrized under $P^C$:
\beq
\frac{\otimes_i\ket{p_i, q_i}+\otimes_i\ket{-p_i, -q_i}}{\sqrt{2}},~~\exists~ (p_i,q_i)\neq (0,0).
\eeq
These together with $\otimes_i\ket{p_i=0,q_i=0}$ form the $(n^{2g}+1)/2$ states in the untwisted sector that are invariant under $\mathbb{Z}_2^C$. 

Next we consider the twisted sectors. Since each of these sectors contains the same number of degenerate ground states, without loss of generality we can assume that $a$ and $b$ change sign along only one cycle $\alpha_g$. Continuity of the gauge potential requires that the states in this twisted sector come in form of 
\beq
\left(\bigotimes_{i'=1}^{g-1}\ket{p_{i^\prime}, q_{i^\prime}}\right)\otimes \ket{p_g=0,q_g=0},
\eeq
which can be deduced from a standard method of considering the double cover of $\Sigma_g$ to make $a$ and $b$ single-valued. Symmetrizing $\otimes_i\ket{p_{i^\prime}, q_{i^\prime}}$ under $\mathbb{Z}_2^C$ gives rise to $(n^{2g-2}+1)/2$ inequivalent states, which is the same as the number of states in the untwisted sector on $\Sigma_{g-1}$. 

In total, the ground state degeneracy of the $S[O(2)\times O(2)]$ BF theory with odd $n$ on genus $g$ Riemann surfaces is
\beq
N_g^\mathrm{odd}=\left(2^{2g}-1\right)\frac{n^{2g-2}+1}{2}+\frac{n^{2g}+1}{2},\label{eq:oddSg}
\eeq
which is the same as the ground state degeneracy of $D^\omega(D_n)$ with odd $n$. The quantum dimension of anyons is uniquely determined by $N_g$. Therefore, we have shown that the anyons in the $S[O(2)\times O(2)]$ BF theory with odd $n$ have the same quantum dimension as those in $D^\omega(D_n)$ with odd $n$.

\subsubsection{Even $n$}

The ground state degeneracy of the $S[O(2)\times O(2)]$ BF theory on a genus $g$ Riemann surface $\Sigma_g$ for even $n$ can be calculated in parallel to the odd $n$ case. Note that for even $n$, besides $\ket{0,0}$, the states $$\ket{\frac{n}{2},0},~\ket{k,\frac{n}{2}}~\ket{\frac{n}{2}+k,\frac{n}{2}}$$ are also invariant under $\mathbb{Z}_2^C$. Therefore, upon gauging, the ground states in the untwisted sectors are symmetrized as 
\bea\label{eq:evenstate}
\begin{aligned}
\frac{\otimes_j\ket{p_j, q_j}+\otimes_j\ket{-p_j, -q_j}}{\sqrt{2}},&~~~\text{if}~\exists~ (p_j,q_j)\notin I,\\
\text{ and }\otimes_i\ket{p_j, q_j}, &~~~\text{if}~ \forall j,~(p_j,q_j)\in I,
\end{aligned}
\eea
where we defined the set \begin{equation}\label{setI}
I\equiv\{(0,0),(0,\frac{n}{2}),(\frac{n}{2}+k,0),(\frac{n}{2}+k,n)\}.
\end{equation}
The total number of such ground states is $(n^{2g}+2^{2g})/2$. In the twisted sector with a twist along the cycle $\alpha_g$, the states subject to the continuity of gauge fields $a$ and $b$ read
\beq
\ket{p_g,q_g}\otimes\ket{p,q}_{g-1},~(p_g,q_g)\in I,
\eeq
where $\ket{p,q}_{g-1}$ is a shorthand for the symmetrized states in the untwisted sector on a genus $(g-1)$ Riemann surface $\Sigma_{g-1}$, given in \Eq{eq:evenstate}. Consequently, the total number of the ground states in this twisted sector is $$4\times\frac{1}{2}\left(n^{2g-2}+2^{2g-2}\right).$$ Since each twisted sector has the same ground state degeneracy, the total ground state degeneracy of the $S[O(2)\times O(2)]$ BF theory with even $n$ on $\Sigma_g$ is
\beq
N_g^\mathrm{even}=2\left(2^{2g}-1\right)\left(n^{2g-2}+2^{2g-2}\right)+\frac{n^{2g}+2^{2g}}{2},~~~\label{evenSg}
\eeq
which is consistent with the ground state degeneracy of $D^\omega(D_n)$ with even $n$. As a result, the anyon quantum dimensions in the two theories also match.

%The ground state degeneracy of the $S[O(2)\times O(2)]$ BF theory with even or odd $n$ can be written in a uniform way
%\beq
%N_g=r\left(2^{2g}-1\right)\frac{n^{2g-2}+r^{2g-2}}{2}+\frac{n^{2g}+r^{2g}}{2},
%\eeq
%with $r=\frac{1}{2}(3+(-1)^n)$.

\subsection{Topological Spins and Fusion Rules }

In this section we present the study of topological spins and fusion rules of anyons via a generalization of the results for the $U(1)\times U(1)$ BF theory and an explicit calculation of modular $S$ and $T$ matrices on a torus with $g=1$. The $S$ and $T$ matrices can fully establish the correspondence between twisted dihedral quantum double and $S[O(2)\times O(2)]$ BF theory.

\subsubsection{Odd $n$}

In the main text we have shown the topological statistics of Wilson loops, and topological spin of twist operators. Here we present a full analysis of the $S$ and $T$ transformations. These transformations map the untwisted sector to itself, but generally map one twisted sector to another twisted sector. More concretely, it yields
\bea
&\hat{S}:~\mathcal{H}_{\pm\pm}\rightarrow\mathcal{H}_{\pm\pm},~\mathcal{H}_{\pm\mp}\rightarrow\mathcal{H}_{\mp\pm},\\
&\hat{T}:~\mathcal{H}_{+\pm}\rightarrow\mathcal{H}_{+\pm},~\mathcal{H}_{-\pm}\rightarrow\mathcal{H}_{-\mp},
\eea
where $\mathcal{H}_{\bullet\bullet}$ are the Hilbert subspaces of the sectors, whose first and second subscripts denote the twist across the holonomy $\alpha$ and $\beta$, respectively, with $-$($+$) signals a twist (untwist). 

For a $U(1)\times U(1)$ BF theory, which is Abelian and does not possess twisted sectors, the $S$ and $T$ transformations act on the states as
% \cite{Esko1993,Liu2013,Mong2022}
\beq
&&\hat{S}\ket{p,q}=\frac{1}{n}\sum_{p^\prime,q^\prime}\exp{\left(2\pi i\frac{n(pq^\prime+qp^\prime)-2kqq^\prime}{n^2}\right)}\ket{p^\prime,q^\prime},\\
&&\hat{T}\ket{p,q}=\exp{\left(2\pi i\frac{q(np-kq)}{n^2}\right)}\ket{p,q}.
\eeq
These relations revealing the mutual and self statistics of the Abelian anyons and can be obtained from the Verlinde formula
% \cite{Verlinde1988,Dijkgraaf1989,CFT} 
and topological spins of the Abelian anyons.
For states in the untwisted sector $\mathcal{H}_{++}$ of the $S[O(2)\times O(2)]$ BF theory created by Wilson loops $W_{p,q}$, the $S$ and $T$ transformations read
\beq
&&\hat{S}\ket{p,q}_{++}=\frac{2}{n}\sum_{(p^\prime,q^\prime)\neq (0,0)}\cos{\left(2\pi \frac{n(pq^\prime+qp^\prime)-2kqq^\prime}{n^2}\right)}\ket{p^\prime,q^\prime}_{++}+\frac{\sqrt{2}}{n}\ket{0,0}_{++},\quad (p,q)\neq (0,0),\\
&&\hat{S}\ket{0,0}_{++}=\frac{1}{n}\ket{0,0}_{++}+\frac{\sqrt{2}}{n}\sum_{(p^\prime, q^\prime)\neq (0,0)}\ket{p^\prime,q^\prime}_{++},\\
&&\hat{T}\ket{p,q}_{++}=\exp{\left(2\pi i\frac{q(np-kq)}{n^2}\right)}\ket{p,q}_{++},\label{eq:STpp}
\eeq
where the $S[O(2)\times O(2)]$ states (with subscripts) and the $U(1)\times U(1)$ states (without subscripts) are related by
\beq
\ket{p,q}_{++}=\frac{1}{\sqrt{2}}\left(\ket{p,q}+\ket{-p,-q}\right)
\eeq
for $(p,q)\neq (0,0)$, and $\ket{0,0}_{++}=\ket{0,0}$.
% \begin{figure*}[t]
%   \centering
% 		\includegraphics[width=0.95\linewidth]{amb}
% 		\caption{$\mathbb{Z}_2$ sign of the $S$ and $T$ transformations.}\label{fig:ambiguity}
% \end{figure*}
And for states in twisted sectors
\bea
&\hat{S}\ket{0,0}_{\pm\mp}=\ket{0,0}_{\mp\pm},~\hat{S}\ket{0,0}_{--}=(-)^\nu\ket{0,0}_{--},~\\
&\hat{T}\ket{0,0}_{\pm\mp}=\ket{0,0}_{\pm -},~\hat{T}\ket{0,0}_{--}=(-)^\nu \ket{0,0}_{-+}.
\eea
% This $\mathbb{Z}_2$ sign originates from whether the $\mathbb{Z}_2^C$ part of the gauge structure has a DWT ($\nu=1$) or not ($\nu=0$). Physically, this $\mathbb{Z}_2^C$ DWT corresponds to the gluing of a $\mathbb{Z}_2^C$ SPT phase,
% \cite{Barkeshli2013,Chen2017,Ryu2017,BBCW}, and is reflected in the non-trivial $F$-symbols compared with the topological order without such a DWT. To see its effect, 
Define
\beq
&&\ket{1}=\frac{1}{\sqrt{2}}\left(\ket{0,0}_{++}+\ket{0,0}_{+-}\right),\\
&&\ket{\eta}=\frac{1}{\sqrt{2}}\left(\ket{0,0}_{++}-\ket{0,0}_{+-}\right),\\
&&\ket{\sigma_\pm}=\frac{1}{\sqrt{2}}\left(\ket{0,0}_{-+}\pm(-i)^\nu\ket{0,0}_{--}\right).
\eeq
as those in the main text. Then the $S$ transformation reads
\beq
&&\hat{S}\ket{1}=\frac{1}{2n}\ket{1}+\frac{1}{2n}\ket{\eta}+\frac{1}{2}\ket{\sigma_+}+\frac{1}{2}\ket{\sigma_-}+\frac{1}{n}\sum_{(p^\prime, q^\prime)\neq (0,0)}\ket{p^\prime,q^\prime}_{++},\\
&&\hat{S}\ket{\eta}=\frac{1}{2n}\ket{1}+\frac{1}{2n}\ket{\eta}-\frac{1}{2}\ket{\sigma_+}-\frac{1}{2}\ket{\sigma_-}+\frac{1}{n}\sum_{(p^\prime, q^\prime)\neq (0,0)}\ket{p^\prime,q^\prime}_{++},\\
&&\hat{S}\ket{\sigma_\pm}=\frac{1}{2}\left(\ket{1}-\ket{\eta}\pm(-)^\nu\ket{\sigma_+}\mp(-)^\nu\ket{\sigma_-}\right),
\eeq
and the $T$ transformation reads
\beq
\hat{T}\ket{1}=\ket{1},~\hat{T}\ket{\eta}=\ket{\eta},~\hat{T}\ket{\sigma_\pm}=\pm i^\nu\ket{\sigma_\pm},
\eeq
which agrees with the direct calculation of $S$ and $T$ matrices of the quantum double $D^{[2k+\nu]}(D_n)$ using finite group data in Sec. SI. In following we present a simple example, $n=1$, to illustrate the effect of $\mathbb{Z}_2$ sign $\nu$.

For $n=1$, there is no even DWT, and the $U(1)\times U(1)$ BF theory describes a phase without topological order. Therefore, gauging the $\mathbb{Z}_2^C$ symmetry results in the $D(\mathbb{Z}_2)$ phase, where the $\psi$ anyons do not exist. Based on examining the fusion rule and topological spin, we identify the $1,\eta,\sigma_+,\sigma_-$ anyons with the $1,e,m,f$ anyons in the toric code model. Although even DWT is absent, the $\mathbb{Z}_2$ toric code phase does admit an odd DWT: this is the double semion model, which can be constructed by gluing a $\mathbb{Z}_2$ SPT phase to the toric code model. In the double semion model, the fusion rule of the anyons remain the same as the toric code model; however, the topological spins of $m$ and $f$ anyons become 1/4 and 3/4, respectively. As expected, the odd DWT comes from the self DWT of the $\mathbb{Z}_2^C$ part of the gauge structure, instead of the (trivial) $U(1)\times U(1)$ part. Thus it is reasonable to expect that for general odd $n$, the topological properties of the anyons that are solely inherited from the $U(1)\times U(1)$ BF theory, specifically the topological spins of the $\psi$ anyons, should not be alternated by the odd DWT in the $\mathbb{Z}_2^C$ part of the gauge structure. On the other hand, the topological properties of the anyons that are arising from the $\mathbb{Z}_2^C$ gauge structure, namely the topological spins and the $S$ transformations of the $\sigma_\pm$ anyons, should only depend on the parity of the DWT. Indeed, the $\psi$ anyons in $D^{[2k]}(D_n)$ have the same topological spins as those in $D^{[2k+1]}(D_n)$. In addition, for all $k$ and odd $n$, the $\sigma_+$ anyon has a topological spin 0 in $D^{[2k]}(D_n)$ and $1/4$ in $D^{[2k+1]}(D_n)$, and the $\sigma_-$ anyon has topological spin $1/2$ in $D^{[2k]}(D_n)$ and $3/4$ in $D^{[2k+1]}(D_n)$, respectively. The $S$ matrix in the subspace spanned by $\sigma_\pm$ reads
\beq
S_{\sigma}=\pm\frac{1}{2}\begin{pmatrix}
1 & -1\\ -1 & 1
\end{pmatrix},
\eeq
with the plus (minus) sign for even (odd) DWTs, which is independent of $n$ and $k$. Therefore, we conclude that in the $S[O(2)\times O(2)]$ BF theory, while \emph{even} DWTs are explicitly reflected in the self CS term of level-$2k$, \emph{odd} DWTs are given implicitly by the DWT in the $\mathbb{Z}_2^C$ gauge structure. This is consistent with the explicit calculation of $S$ and $T$ matrices using finite group modular data in the Sec. SI.

\subsubsection{Even $n$}

\begin{table}[!htbp]
\begin{tabular}{cccccc}
\hline\hline
Anyon & Quantum & Number &Conjugacy & Centra- & \multirow{2}{*}{Irrep.} \\
Label&Dim.&of Anyons& Class & lizer &\\
\hline
$1_\mu$ & $1$ & $4$ & $\{1\}$ & $D_n$ & $\mathbf{1}_\mu$\\
$e_\mu$ & $1$ & $4$ & $\{C^{n/2}_n \}$ & $D_n$ & $\mathbf{1}_\mu$\\
$\psi_l$ & $2$ & $(n-2)/2$ & $\{1\}$ & $D_n$ & $\mathbf{2}_l$\\
$\bar{\psi}_l$ & $2$ & $(n-2)/2$ & $\{C^{n/2}_n\}$ & $D_n$ & $\mathbf{2}_l$\\
$\psi_{l,r}$ & $2$ & $n(n-2)/2$ & $\{C_n^l,C_n^{n-l}\}$ & $\mathbb{Z}_n$ & $\mathbf{1}_r$\\
$m_\mu$ & $n/2$ & $4$ & $\{M_1,...,M_n\}$ & $D_2$ & $\mathbf{1}_\mu$\\
$f_\mu$ & $n/2$ & $4$ & $\{C_nM_1,...,C_nM_n\}$ & $D_2$ & $\mathbf{1}_\mu$\\
\hline\hline
\end{tabular}
\caption{Anyon data of $D^\omega(D_n)$ for even $n$. In the first and the last two rows $\mu=1,e,m,f$ label the four different one dimensional irreps of $D_n$ ($D_2=\mathbb{Z}_2\times\mathbb{Z}_2$). The direct products of these irreps obey the the same fusion rules as the 1, $e$, $m$ and $f$ anyons in $\mathbb{Z}_2$ toric code model. For simplicity the trivial anyon $1_1$ will be also denoted as 1. There are $(n^2+28)/2$ distinct anyons, and the total quantum dimension is $4n^2$. Here $l=1,...,(n-2)/2$ and $r=-(n-2)/2,...,(n-2)/2$. 
}\label{table:even}
\end{table}

Similar to the odd $n$ case, the $\psi$ anyons with quantum dimension 2 are identified with the Wilson loop operators of the $S[O(2)\times O(2)]$ BF theory
\beq
W_{p,q}=2\mathcal{P}\cos\left(\oint_\gamma pa+qb\right),~(p,q)\notin I.
\eeq
Their fusion rule, consistent with those of $\psi$ anyons in $D^\omega(D_n)$, is the same as \Eq{eq:wfusionodd}. To be concrete, here $W_{p,0}$ and $W_{p,\frac{n}{2}}$ are identified as $\psi_l$ and $\bar{\psi}_l$, and the other Wilson loop operators are identified with $\psi_{l,r}$. Aside from the Wilson loops and the identity operator, there are 15 twist operators, among which 7 have quantum dimension 1 and 8 have quantum dimension $n/2$. Their topological spin and fusion rule can be determined from the $S$ and $T$ matrices, similar to the odd $n$ case. In the untwisted sector (labeled by $++$), the $S$ and $T$ transformations constrained to the basis of states $$\ket{++}=\Big(\ket{0,0}_{++},\ket{\frac{n}{2},0}_{++},\ket{k,\frac{n}{2}}_{++},\ket{\frac{n}{2}+k,\frac{n}{2}}_{++}\Big)^\mathbf{T}$$ are given by $\hat{S}\ket{++}=\frac{1}{n}S_0\ket{++}$, $\hat{T}\ket{++}=T_0\ket{++}$, where the matrices $S_0$ and $T_0$ read
\beq
S_0&=&\begin{pmatrix}
1 & 1 & 1 & 1\\
1 & 1 & (-)^\frac{n}{2} & (-)^\frac{n}{2}\\
1 & (-)^\frac{n}{2} & (-)^k & (-)^{\frac{n}{2}+k}\\
1 & (-)^\frac{n}{2} & (-)^{\frac{n}{2}+k} & (-)^k
\end{pmatrix},\\
T_0&=&\mathrm{diag}\{1,1,i^k,i^{n+k}\}.
\eeq
Apparently the $S$ transformation will also generate states $\ket{p,q}_{++}$ with $(p,q)\notin I$, which are the same as \Eq{eq:STpp} and omitted for clarity. For twisted sectors, the actions of $S$ and $T$ still have a $\mathbb{Z}_2$ sign arising from whether $\mathbb{Z}_2^C$ has nontrivial DWT or not, similar to the odd DWTs in the odd $n$ case. After considering this $\mathbb{Z}_2$ sign labeled by $\nu=0,1$, the $S$ and $T$ transformations are
\bea
&\hat{S}\ket{\pm\mp}=\frac{1}{2}S_0\ket{\mp\pm},\quad\hat{S}\ket{--}=\frac{1}{2}(-)^\nu S_0\ket{--},\\
&\hat{T}\ket{\pm\mp}=T_0\ket{\pm -},\quad \hat{T}\ket{--}=(-)^\nu T_0\ket{-+}.
\eea
Here the basis of states $\ket{\bullet\bullet}$ is defined similarly to $\ket{++}$ as $$\ket{\bullet\bullet}=\Big(\ket{0,0}_{\bullet\bullet},\ket{\frac{n}{2},0}_{\bullet\bullet},\ket{k,\frac{n}{2}}_{\bullet\bullet},\ket{\frac{n}{2}+k,\frac{n}{2}}_{\bullet\bullet}\Big)^\mathbf{T}.$$ The 1/2 factor in the $S$ transformation comes from the total number of states equal to $4=2^2$ in each twisted sector, similar to the $1/n$ factor in the untwisted sector [see \Eq{eq:STpp}].

However, for even $n$, the dihedral group $D_n$ has a non-trivial center $\mathbb{Z}_2^\pi=\{1,C_n^{n/2}\}$ as shown in Table \ref{table:even}, which commutes with all group elements including the $\mathbb{Z}_2^C$ action. Thus, it is possible that this center $\mathbb{Z}_2^\pi$ has a nontrivial mixed DWT with $\mathbb{Z}_2^C$. This mixed DWT is independent of the self DWT of $\mathbb{Z}_2^C$ (labeled by $\nu$) or $\mathbb{Z}_n\supset\mathbb{Z}_2^\pi$ (labeled by $k$), and it incarnates in the $S[O(2)\times O(2)]$ BF theory as an additional $\mathbb{Z}_2$ sign labeled by $\upsilon=0,1$ in the $S$ and $T$ transformations in the twisted sectors:
\bea
&\hat{S}\ket{\pm\mp}=\frac{1}{2}S_0\ket{\mp\pm},~ \hat{S}\ket{--}= \frac{1}{2}(-)^\nu S_0\Gamma^2_{\upsilon}\ket{--},~~~\\
&\hat{T}\ket{\pm\mp}=T_0\ket{\pm -},~ \hat{T}\ket{--}=(-)^\nu T_0\Gamma_{\upsilon}^2\ket{-+},
\eea
where $\Gamma_{\upsilon}=\mathrm{diag}\{1,1, i^{\upsilon}, i^{\upsilon}\}$.
Heuristically, while the $\mathbb{Z}_2^C$ self DWT flips the sign of all the four states in the $--$ sector in an $S$ or $T$ transformation, the mixed DWT only flips the sign of the two states $\ket{k,\frac{n}{2}}_{--}$ and $\ket{\frac{n}{2}+k,\frac{n}{2}}_{--}$ dependent on the self DWT in $\mathbb{Z}_2^\pi$, reflecting its mixed nature. Note that the self DWT in $\mathbb{Z}_2^\pi$ is exactly encoded in the level-$2k$ self CS term of the $S[O(2)\times O(2)]$ BF theory, and hence relies on $k$, since $\mathbb{Z}_2^\pi$ is a subgroup of $\mathbb{Z}_n$.

In parallel to the odd $n$ case, we define
\begin{subequations}\label{eq:eventwist}
\begin{align}
&\ket{\mathbf{1}}=(\ket{1},\ket{1_e},\ket{1_m},\ket{1_f})^\mathbf{T}=\frac{1}{\sqrt{2}}\left(\ket{++}+\Gamma_{\upsilon}^{-1}\ket{+-}\right),\\
&\ket{\mathbf{e}}=(\ket{e_1},\ket{e_e},\ket{e_m},\ket{e_f})^\mathbf{T}=\frac{1}{\sqrt{2}}\left(\ket{++}-\Gamma_{\upsilon}^{-1}\ket{+-}\right),\\
&\ket{\mathbf{m}}=(\ket{m_1},\ket{m_e},\ket{m_m},\ket{m_f})^\mathbf{T}=\frac{1}{\sqrt{2}}\left(\ket{-+}+(-i)^\nu\Gamma_{\upsilon}^{-1}\ket{--}\right),\\
&\ket{\mathbf{f}}=(\ket{f_1},\ket{f_e},\ket{f_m},\ket{f_f})^\mathbf{T}=\frac{1}{\sqrt{2}}\left(\ket{-+}-(-i)^\nu\Gamma_{\upsilon}^{-1}\ket{--}\right).
\end{align}
\end{subequations}
Then the $S$ and $T$ transformations in the 16-dimensional subspace spanned by basis \Eq{eq:eventwist} read
\beq
&&S_\mu =\frac{1}{4n}\begin{pmatrix}
2S_0 & 2S_0 & nS_{\upsilon}& nS_{\upsilon}\\
2S_0 & 2S_0 & -nS_{\upsilon} & -nS_{\upsilon}\\
nS_{\upsilon}^\mathbf{T} & -nS_{\upsilon}^\mathbf{T} & nS_{\nu,\upsilon} & -nS_{\nu,\upsilon}\\
nS_{\upsilon}^\mathbf{T} & -nS_{\upsilon}^\mathbf{T} & -nS_{\nu,\upsilon} & nS_{\nu,\upsilon}
\end{pmatrix},\label{eq:evenS}\\
&&T_\mu =\mathrm{diag}\{T_0,T_0,i^\nu\Gamma_{\upsilon} T_0,-i^\nu\Gamma_{\upsilon} T_0\},\label{eq:evenT}
\eeq
where $S_{\upsilon}=\Gamma_{\upsilon}^{-1}S_0$ and $S_{\nu,\upsilon}=(-)^\nu\Gamma_{\upsilon}^{-1}S_0\Gamma_{\upsilon}^{-1}$. The quantum dimensions of the twist operators can be read out from the first row of the $S$ matrix, yielding that $1_\mu$ and $e_\mu$ anyons have quantum dimension 1, while $m_\mu$ and $f_\mu$ anyons have quantum dimension $n/2$, for $\mu=1,e,m,f$. These are consistent with Table \ref{table:even}. In conclusion, all the $4n$ DWTs of $D^\omega(D_n)$ for even $n$ are captured by $\omega=[k,\nu,\upsilon]$. In the following section we will illustrate the physics of anomalous DWT, especially the mixed one, through a simple example of $D_2$.

Unlike the $n>2$ case where $D_n$ is a semidirect product of $\mathbb{Z}_n$ and $\mathbb{Z}_2$, the $n=2$ case, $D_2$, has a direct product structure: $D_2=\mathbb{Z}_2^\pi\times\mathbb{Z}_2^C$. As a result, there are two ways to describe a $D^\omega(D_2)$ phase. One is the $S[O(2)\times O(2)]$ BF theory proposed in this work
\beq
\mathcal{L}_{D_2}=\frac{ik}{2\pi}a\mathrm{d}a+\frac{i}{\pi}a\mathrm{d}b,
\eeq
with $k=0,1$ labeling the \emph{regular} DWTs. Physically, this can be viewed as a toric code (for $k=0$) or double semion (for $k=1$) model with a $\mathbb{Z}^C_2$ orbifold. In this representation, the \emph{anomalous} DWTs are implicitly captured by the $S$ and $T$ transformations on states in the twisted sectors originated from the orbifold. Since $\mathbb{Z}_2^\pi$ and $\mathbb{Z}_2^C$ commute with each other, this orbifold procedure can be equivalently described by stacking, which gives the other way to describe $D^{[k,\nu,\upsilon]}(D_2)$. In this representation, the quantum double phase is viewed as a stacking of two layers of $\mathbb{Z}_2$ quantum double phases
\beq\label{eq:stack}
\mathcal{L}_{D_2}^\prime =\frac{ik}{2\pi}\alpha\mathrm{d}\alpha+\frac{i}{\pi}\alpha\mathrm{d}\beta+\frac{i\nu}{2\pi}\alpha^\prime\mathrm{d}\alpha^\prime+\frac{i}{\pi}\alpha^\prime\mathrm{d}\beta^\prime+\frac{i\upsilon}{2\pi}\alpha\mathrm{d}\alpha^\prime,
\eeq
here $\alpha$ and $\alpha^\prime$ are $\mathbb{Z}_2^\pi$ and $\mathbb{Z}_2^C$ gauge fields, with $\beta$ and $\beta^\prime$ being the Lagrangian multipliers, respectively. A Maxwell term also exists for the field $\alpha$ and $\alpha^\prime$. As in the $S[O(2)\times O(2)]$ BF theory, $k=0,1$ captures the regular DWTs arising from the self DWT of $\mathbb{Z}_2^\pi$; however, the anomalous DWTs from $\mathbb{Z}_2^C$ is also explicitly captured by the level-$2\nu$ self CS term in $\alpha^\prime$. In addition, the mixed DWT between the two $\mathbb{Z}_2$ gauge structures, which corresponds to a mutual CS term between the two $\mathbb{Z}_2$ gauge fields $\alpha$ and $\alpha^\prime$, is explicitly captured by the $\upsilon$-term in the stacking theory \Eq{eq:stack}. Thus, compared with the $S[O(2)\times O(2)]$ BF theory, the stacking theory \Eq{eq:stack} reveals all the $2^3=8$ possible DWTs. 

The anyons in the stacking theory are bound states of two anyons from each $\mathbb{Z}_2$ quantum double layer. We use $\mu_\mu^\prime$ with $\mu,\mu^\prime=1,e,m,f$ to label these anyons, where $e$ and $m$ anyons carrying ``electric" and ``magnetic" charges are coupled to $\alpha,\alpha^\prime$ and $\beta,\beta^\prime$, respectively. Under this basis, the $S$ and $T$ matrices of the stacking theory \Eq{eq:stack} are equal to those of the $S[O(2)\times O(2)]$ BF theory calculated in the previous subsection, \Eq{eq:evenS} and \Eq{eq:evenT}, for all $(k,\nu,\upsilon)$. This strongly corroborates the $S[O(2)\times O(2)]$ BF theory as an effective theory of the $D^{[k,\nu,\upsilon]}(D_n)$ phase for even $n$.

The effects of DWTs, which are rather abstract in the $S[O(2)\times O(2)]$ BF theory, can be easily read out from the anyon statistics in the stacking theory \Eq{eq:stack}. Consider the intra-layer anyons in the $D(\mathbb{Z}_2^\pi)$ and the $D(\mathbb{Z}_2^C)$ layers (anyons bound with the trivial anyon in the other layer), defined as $1,e_1,m_1,f_1$ and $1,1_e,1_m,1_f$, respectively. Here 1 is the bound state of two trivial anyons in each layer, which becomes the trivial anyon in the stacking theory. Without DWT, both of the two layers are in the $\mathbb{Z}_2$ toric code phase. A regular DWT in the $D(\mathbb{Z}_2^\pi)$ layer with $k=1$ in \Eq{eq:stack} converts the layer to be in a double semion phase, while the $D(\mathbb{Z}_2^C)$ layer remains toric code. However, an anomalous DWT in the $D(\mathbb{Z}_2^C)$ will also be resulted in a stacking of a toric code phase and a double semion phase, which is equivalent to the effect of the regular DWT. This occasional isomorphism is originated from the duality between the $\mathbb{Z}_2^C$ and $\mathbb{Z}_n=\mathbb{Z}_2^\pi$ subgroups in $D_{n=2}$, and not expected to exist for general $n>2$. On the other hand, the mixed DWT between $\mathbb{Z}_2^C$ and $\mathbb{Z}_2^\pi$ does not alternate the statistics among the intra-layer anyons; however, the statistics among anyons from different layers will be affected. For example, consider $k=\nu=0$ and $\upsilon=1$ case. Compared with a trivial stacking of two toric code layers, in this phase the statistics among $1_\mu$ and among $\mu_1$ remain the same as the toric code model; however, the mutual statistics between $1_m$ and $m_1$ becomes non-trivial, with a $\pi/2$ mutual statistical angle. Thus, for generic even $n$, we expect an anomalous DWT labeled by $\upsilon$ will not affect the statistics among the $1_\mu$ anyons and among the $\mu_1$ anyons, but the mutual statistics between these two groups.

\section{Monte Carlo Simulation with Monopole Suppression}

In this section we discuss the procedure of suppressing monopoles in the simulation, and show the obtained phase diagram. The real $O(2)$ gauge field $\mathbf{u}_{\mathbf{i}\mathbf{j}}$ can be parameterized by its $U(1)$ part $a_{\mathbf{i}\mathbf{j}}$ and $\mathbb{Z}_2$ part $\sigma_{\mathbf{i}\mathbf{j}}$ as
\beq
\mathbf{u}_{\mathbf{i}\mathbf{j}}=\mathbf{u}_{\mathbf{j}\mathbf{i}}^\mathbf{T}=\begin{pmatrix}
    \cos{n a_{\mathbf{i}\mathbf{j}}} & \sin{n a_{\mathbf{i}\mathbf{j}}} \\
    -\sigma_{\mathbf{i}\mathbf{j}}\sin{ na_{\mathbf{i}\mathbf{j}}} & \sigma_{\mathbf{i}\mathbf{j}}\cos{ na_{\mathbf{i}\mathbf{j}}}
\end{pmatrix}=\mathbf{Z}^{\frac{1+\sigma_{\mathbf{i}\mathbf{j}}}{2}}\cdot\left(\mathbf{I}\cos{n a_{\mathbf{i}\mathbf{j}}}+i\mathbf{Y}\sin{n a_{\mathbf{i}\mathbf{j}}}\right),
\eeq
where $a_{\mathbf{i}\mathbf{j}}\in [0,2\pi)$ and $\sigma_{\mathbf{i}\mathbf{j}}=\pm 1$. The Maxwell coupling on a plaquette can be rewritten as
\beq
H_\square =-\frac{1}{2\kappa}\mathrm{tr}~\mathbf{u}_{12}\cdot\mathbf{u}_{23}\cdot\mathbf{u}_{34}\cdot\mathbf{u}_{41}
=-\frac{1+\sigma_{12}\sigma_{23}\sigma_{34}\sigma_{41}}{\kappa}\cos{n \Psi_\square},
\eeq
where $\Psi_\square=a_{12}\sigma_{23}\sigma_{34}\sigma_{41}+a_{23}\sigma_{34}\sigma_{41}+a_{34}\sigma_{41}+a_{41}$. Denote $K_\square=(1+\sigma_{12}\sigma_{23}\sigma_{34}\sigma_{41})/(\kappa T)$ where $T$ is the temperature. Then the Boltzmann weight can be approximated by means of the Villain approximation
\beq
e^{-\beta H_\square}=\sum_{r_{\bar{\mathbf{i}}\bar{\mathbf{j}}}=-\infty}^{+\infty}\exp\left(-\frac{K_\square}{2}\left(n \Psi_\square-2\pi r_{\bar{\mathbf{i}}\bar{\mathbf{j}}}\right)^2\right),
\eeq
where $\left<\bar{\mathbf{i}}\bar{\mathbf{j}}\right>$ is the dual lattice link piercing the plaquette $\square$. The monopole suppression condition under the Villain approximation is $\nabla\cdot \mathbf{r}_{\bar{\mathbf{i}}}=0$ for $\mathbf{r}_{\bar{\mathbf{i}}}=(r_{\bar{\mathbf{i}},\bar{\mathbf{i}}+\hat{\mathbf{x}}},r_{\bar{\mathbf{i}},\bar{\mathbf{i}}+\hat{\mathbf{y}}},r_{\bar{\mathbf{i}},\bar{\mathbf{i}}+\hat{\mathbf{z}}})$. It can be enforced in the partition function by introducing the lagrange multiplier $\gamma_{\bar{\mathbf{i}}}\in\mathbb{R}$ defined on the dual site $\bar{\mathbf{i}}$. The partition function for the gauge field with monopole suppression then reads
\beq\label{Z_monopole_suppression}
Z=\sum_{\{\sigma=\pm 1\}}\sum_{\{r\in\mathbb{Z}\}}\int_{-\infty}^{+\infty}D[a]\int_{-2\pi}^{2\pi}D[\gamma]\prod_{\bar{\mathbf{i}}}\exp\left(i\gamma_{\bar{\mathbf{i}}}\nabla\cdot\mathbf{r}_{\bar{\mathbf{i}}}\right)\prod_\square\exp\left(-\frac{K_\square [a,\sigma]}{2}\left(n \Psi_\square [a,\sigma]-2\pi r_{\bar{\mathbf{i}}\bar{\mathbf{j}}}\right)^2\right).
\eeq
Note that $a_{\mathbf{i}\mathbf{j}}$ is now defined through $\mathbb{R}$ due to the Villain approximation. Its periodicity is recovered by the integer field $r_{\bar{\mathbf{i}}\bar{\mathbf{j}}}$.

A more rigorous argument about the stability of the multicritical point is given as follows. First consider a three-dimensional fixed volume $\tilde{\mathcal{V}}$ (preserved under the RG flow) as a subspace of the five-dimensional phase diagram of the effective field theory \eqref{eq:WF} in the main text. For such a $\tilde{\mathcal{V}}$ with $r_v$ and $r_s$ remaining irrelevant in it and containing the $O(2)\times O(1)$ symmetric fixed point, any RG trajectory---as long as it is not the direct trajectory from $O(2)\times O(1)$ to the $O(3)$ symmetric fixed point---will be driven away from the latter, since $t_v$ and $t_s$ are always relevant in the entire parameter space. This suggests the stability of the multicritical point within $\tilde{\mathcal{V}}$. On the other hand, the phase diagram $\mathcal{V}$ of the lattice model \eqref{eq:LH} in the main text spanned by $J_v$, $J_s$, and $\kappa$ is also an embedded three-dimensional subspace. \emph{Assuming} that $r_v$ and $r_s$ are always irrelevant in $\mathcal{V}$, the lattice model phase diagram $\mathcal{V}$ can be deformed to the phase diagram in $\tilde{\mathcal{V}}$ with irrelevant $r_{v,s}$. This assumption is justified by the observation that upon varying $J_s$ and $J_v$ in the lattice model \eqref{eq:LH}, the universality class of transitions between stable phases remains Ising or $U(1)$, where the $r_v$ and $r_s$ are indeed irrelevant. Thus, the multicritical point should also be stable in the lattice model.

\section{The Complex Lattice Model}

In this section we present another lattice model involving complex boson fields, where the $O(2)$ gauge group manifests as $O(2,\mathbb{C})$. Consider the classical theory defined on 3D cubic lattice
\beq
H=H_\mathrm{Maxwell}+\sum_{\left<\mathbf{i}\mathbf{j}\right>}-J_\Phi\mathbf{\Phi}_\mathbf{i}^*\cdot\mathbf{u}_{\mathbf{i}\mathbf{j}}\cdot\mathbf{\Phi}_\mathbf{j}-J_\phi \phi_\mathbf{i}|\mathbf{u}_{\mathbf{i}\mathbf{j}}|\phi_\mathbf{j}+\sum_\mathbf{i}\lambda \phi_\mathbf{i}\left(|\Phi_{1,\mathbf{i}}|^2-|\Phi_{2,\mathbf{i}}|^2\right)+\Lambda|\Phi_{1,\mathbf{i}}-\Phi^*_{2,\mathbf{i}}|^2,
\label{eq:CL}
\eeq
where $\mathbf{\Phi}_\mathbf{i}=(\Phi_{1,\mathbf{i}},\Phi_{2,\mathbf{i}})$ is a unit-norm two-component complex boson field, $\phi_\mathbf{i}=\pm 1$ is an Ising field, and $\mathbf{u}_{\mathbf{i}\mathbf{j}}$ is the $O(2,\mathbb{C})$ gauge field. $J_\Phi$, $J_\phi$, $\lambda$ and $\Lambda$ are coupling constants. Without losing of generality, we assume $\lambda,\Lambda>0$. For the matter fields, the $O(2)$ actions read
\bea
&U(1):~\mathbf{\Phi}\mapsto e^{in\theta \mathbf{Z}}\cdot\mathbf{\Phi},~\phi\mapsto\phi,\\
&\mathbb{Z}_2:~\mathbf{\Phi}\mapsto \mathbf{X}\cdot\mathbf{\Phi},~\phi\mapsto -\phi,
\eea
which is different to the real case. Here $\mathbf{X}$ and $\mathbf{Z}$ represent the Pauli matrices. To illustrate the phase diagram of the complex theory, we adopt a Ginzberg-Landau mean-field treatment based on the continuum version of \Eq{eq:CL}. The corresponding Ginzberg-Landau potential reads
\beq
V[\mathbf{\Phi}, \phi]=t_\Phi |\mathbf{\Phi}|^2+|\mathbf{\Phi}|^4+t_\phi \phi^2+\phi^4%+\tilde{t}|\mathbf{\Phi}|^2\phi^2
+\lambda\phi\left(|\Phi_1 |^2-|\Phi_2 |^2\right)+\Lambda|\Phi_1-\Phi^*_2 |^2,\label{eq:GL}
\eeq
where parameters $t_\Phi$ and $t_\phi$ control the condensation of $\mathbf{\Phi}$ and $\phi$, respectively. The coupling constants of the quartic terms are set to be 1 for clarity. Note that the complex theory is closely related to the real theory. Define
\beq
\Psi_1=\Phi_1+\Phi_2^*,~\Psi_2=\Phi_1^*-\Phi_2,\label{eq:psi2}
\eeq
The $O(2)$ action on $\Psi_1$ is the same as that in the real theory, where $U(1)$ and $\mathbb{Z}_2$ act as charge conservation and charge conjugation symmetries, respectively. Assuming $\Psi_2$ is gapped and decoupled from the low energy sector, the Ginzberg-Landau potential can be rewritten in terms of $\Psi_1$ as
\beq
V[\Psi_1,\phi]=t_\Phi|\Psi_1|^2+|\Psi_1|^4+t_\phi \phi^2+\phi^4,
\eeq
which equals that in the real theory before integrating out the gauge field.

Below we only consider the fully deconfined phases in the complex theory. Similar to the case of the real theory, there are four such phases:
\begin{enumerate}
\item The $O(2)$ Coulomb phase: $\left<\Phi_1\right>=\left<\Phi_2\right>=\left<\phi\right>=0$. The matter fields are complex boson $\mathbf{\Phi}$ carrying $\mathbf{2}_n$ irrep and real boson $\phi$ carrying $\mathbf{1}_-$ irrep.
\item The $U(1)$ Coulomb phase: $\left<\Phi_1\right>=\left<\Phi_2\right>=0$, $\left<\phi\right>\neq 0$. The matter fields are charged-$\pm n$ bosons $\Phi_{1,2}$.
\item The $D(D_n)$ phase: $\left<\Phi_1\right>=\left<\Phi_2^*\right>\neq 0$, $\left<\phi\right>= 0$. The matter field is the $\eta$ ($1_{e}$) anyon when $n$ is odd (even).
\item The $D(\mathbb{Z}_n)$ phase: $\left<\Phi_1\right>\neq 0$, $\left<\Phi_2\right>\neq 0$, $\left<\phi\right>\neq 0$. This is a pure gauge theory phase without matter fields.
\end{enumerate}

\begin{figure}[H]
  \centering
		\includegraphics[width=0.3\linewidth]{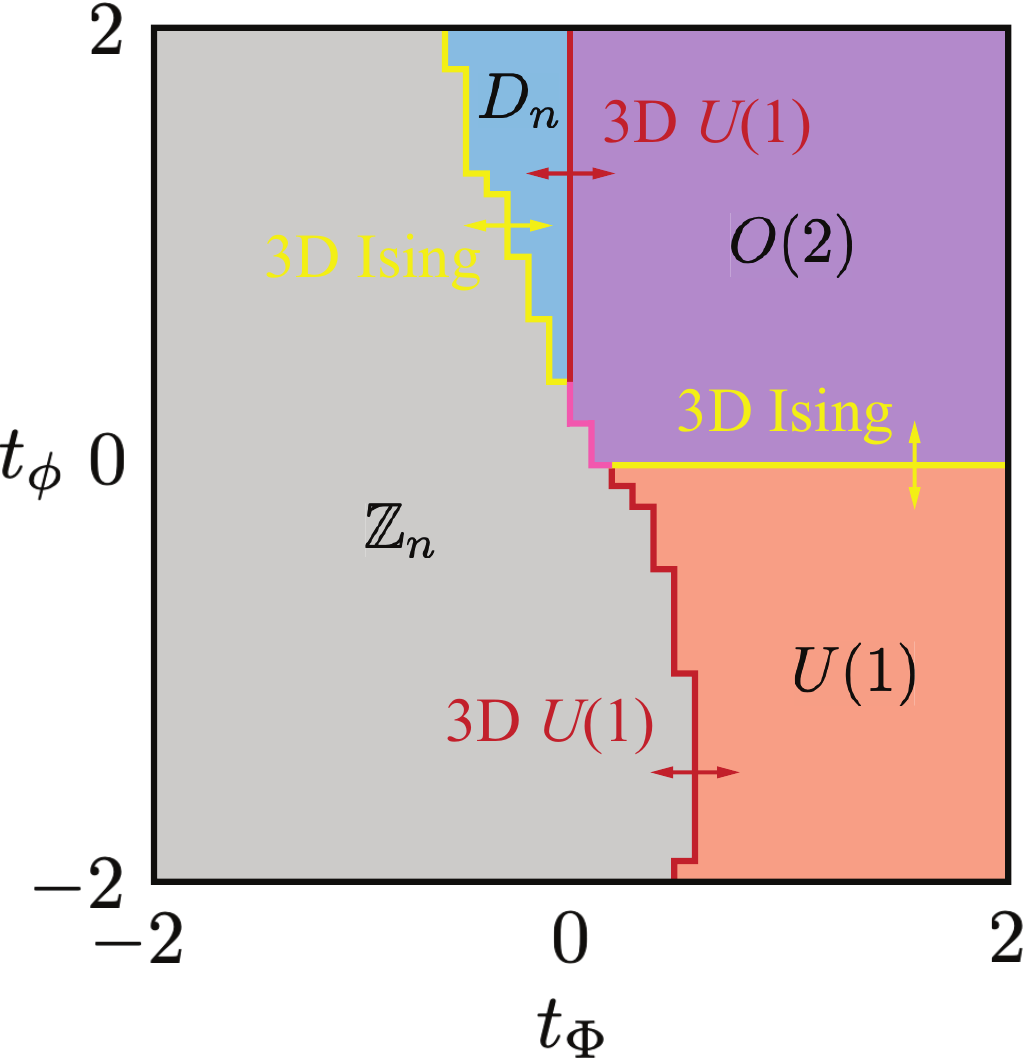}
		\caption{Mean-field phase diagram of the complex $O(2)$ theory \Eq{eq:CL} with $\lambda=1.5$ and $\Lambda=0.2$ when all the possible gauge fields are deconfined. The gauge structure of each phase is shown explicitly. The 3D $U(1)$ and the 3D Ising transitions are marked in red and yellow, respectively. The direct transition between the $O(2)$ Coulomb phase and the $D(\mathbb{Z}_n)$ phase, marked in pink, could be in a coexistence of the 3D $U(1)$ and 3D Ising classes.}\label{fig:MF}
\end{figure}

Assuming that $H_\mathrm{Maxwell}$ deconfines all the possible gauge fields, we calculate the mean-field phase diagram of the complex $O(2)$ theory \Eq{eq:CL} with $\lambda=1.5$ and $\Lambda=0.2$, as shown in FIG. \ref{fig:MF}. The phase transition patterns are almost the same as the real theory case, while the multicritical point with emergent $O(3)$ symmetry is no longer stable due to the relevant $\lambda$-term in \Eq{eq:CL}. When $\Psi_2$ in \Eq{eq:psi2} is decoupled from the low energy sector, the complex theory is reduced to the real theory, eliminating the $\lambda$-term and recovering the stable multicritical point. The $O(2)$ Coulomb phase and the $D(\mathbb{Z}_n)$ phase can still sandwich a direct transition.

\vfill 

\end{document}